\documentclass[prd,,aps,nofootinbib,amsmath,amssymb]{revtex4}
\usepackage{amssymb,amsmath,epsfig,bm}
\usepackage[usenames,dvipsnames]{color}
\usepackage{mathrsfs}
\usepackage[bookmarks]{hyperref}

\newtheorem{theorem}{Theorem}

\newcommand{\Real}{\mathbb{R}}
\newcommand{\sign}{\mbox{sign}}

\begin{document}

\title{Shadow of a naked singularity}
\author{N\'estor Ortiz$^{1}$, Olivier Sarbach$^{1,2}$, and Thomas Zannias$^{2,3}$}

\affiliation{$^1$Perimeter Institute for Theoretical Physics, 31 Caroline Street, Waterloo, Ontario N2L 2Y5, Canada,\\
$^2$Instituto de F\'{\i}sica y Matem\'aticas,
Universidad Michoacana de San Nicol\'as de Hidalgo,\\
Edificio C-3, Ciudad Universitaria, 58040 Morelia, Michoac\'an, M\'exico,\\
$^3$Department of Physics, Queen's University, Kingston, Ontario K7L 3N6, Canada.}
\email{nortiz@perimeterinstitute.ca, sarbach@ifm.umich.mx, zannias@ifm.umich.mx}

\begin{abstract}
We analyze the redshift suffered by photons originating from an external source, traversing a collapsing dust cloud, and finally being received by an asymptotic observer. In addition, we study the shadow that the collapsing cloud casts on the sky of the asymptotic observer. We find that the resulting redshift and properties of the shadow depend crucially on whether the final outcome of the complete gravitational collapse is a black hole or a naked singularity. In the black hole case, the shadow is due to the high redshift acquired by the photons as they approach the event horizon, implying that their energy is gradually redshifted toward zero within a few crossing times associated with the event horizon radius. In contrast to this, a naked singularity not only absorbs photons originating from the source, but it also emits infinitely redshifted photons with and without angular momenta. This emission introduces an abrupt cutoff in the frequency shift of the photons detected in directions close to the radial one, and it is responsible for the shadow masking the source in the naked singularity case. Furthermore, even though the shadow forms and begins to grow immediately after the observer crosses the Cauchy horizon, it takes many more crossing times than in the black hole case for the source to be occulted from the observer's eyes.

We discuss possible implications of our results for testing the weak cosmic censorship hypothesis. Even though at late times the image of the source perceived by the observer looks the same in both cases, the dynamical formation of the shadow and the redshift images has distinct features and time scales in the black hole versus the naked singularity case. For stellar collapse, these time scales seem to be too short to be resolved with existing technology. However, our results may be relevant for the collapse of seeds leading to supermassive black holes.
\end{abstract}

\maketitle

\section{Introduction}

A widely studied collapse model in general relativity is the family of Tolman-Bondi spacetimes, which describe the collapse of a spherical dust cloud. The popularity of this model stems from the fact that in suitable coordinates the metric components and fluid fields can be expressed in terms of elementary functions, and thus  provides exact solutions of the Einstein-Euler equations in the absence of pressure. Curiously, for a large class of initial data satisfying reasonable physical assumptions, these collapse models predict  the formation of shell-focusing singularity a portion of which is null and visible to local observers, see for instance Refs.~\cite{dElS79,dC84,rN86,pJiD93,Joshi-Book,bNfM01,nOoS11}. Furthermore, for suitable initial data in this class, part of this null singularity is visible even to observers in the asymptotic region; that is, the singularity is ``globally naked". While the Tolman-Bondi models, being spherical symmetric and having zero pressure, are highly restrictive from a physical point of view, still it should be emphasized that the occurrence of such globally naked singularities does not require any fine-tuning in the initial data within these models. Although there has been a significant amount of work regarding the properties of such singular spacetimes including the stability of the associated Cauchy horizon~\cite{dC84,bWkL89,iD98,hItHkN98,hItHkN99,hItHkN00,bNtW02,tWbN09,eDbN11b,eDbN11,nOoS14,nOoS15}, relatively little attention has been placed on the manner that a globally naked singularity resulting from the complete gravitational collapse interacts with the rest of the Universe. However, nowadays, technological advances reach to the level where even the minutest predictions of general relativity can be placed under observational scrutiny, the current gravitational wave detectors~\cite{LIGO,VIRGO,KAGRA} and the Event Horizon Telescope project~\cite{EHT} being prime examples that will test the theory in its strong field regime. These developments suggest that it is pertinent to start analyzing observational signatures associated with naked singularities. Even if the cosmic censorship conjecture would  be proven rigorously some time in the future, ultimately its acceptance as a fundamental law of nature would require to be checked in the real Universe. 

In this spirit, in a recent paper~\cite{nOoStZ14} (paper I hereafter), we have proposed  that the asymptotically measured frequency shift suffered by photons traversing a collapsing cloud can serve as a tool capable of differentiating between  the formation of a globally naked singularity and the formation of an event horizon; i.e. this frequency shift can serve as an instrument to test the validity of the weak cosmic censorship hypothesis in terrestrial observatories. Evidence supporting this proposal came from the analysis of radial photons from an external source passing through a collapsing spherical dust cloud. As shown in~\cite{nOoS14b}, the total frequency shift of such photons relative to asymptotic static emitters and observers is always toward the red. Moreover, the analysis in~I revealed that the photon's energy measured by the asymptotic observer is gradually redshifted toward zero in the black hole case, whereas whenever a globally naked singularity forms the redshift of these radial photons remains finite and their frequency shift exhibits an abrupt cutoff once they have been detected by the observer who has crossed the Cauchy horizon (see Figs.~2~and~3~in I).

The main goal of the present  paper is to subject the proposal in~I to further testing by allowing the incidence radiation to have nonvanishing angular momenta. Within the geometric optics approximation, this radiation consists of a collection of photons which are generated at past infinity, propagate freely through the collapsing cloud and are eventually detected by an observer in the asymptotic region. The central focus of this work is to examine the properties of the frequency shift of these photons. As will become clear in the course of this paper, the inclusion of angular momenta has interesting consequences at both the computational and the conceptual levels, and furthermore it adds a new flavor to the proposal in~I. At the computational level, the evaluation of the redshift becomes a more difficult task since it requires integrating nonradial null geodesics inside the collapsing cloud. Nevertheless, as we show in this article, the nonradial photon's frequency shift exhibits a qualitatively similar behavior than in the radial case, provided the direction is close enough to the radial one: first, the frequency shift is always toward the red. Second, this shift gradually converges to zero in the black hole case, whereas there is an abrupt cutoff in the naked singularity case. At the conceptual level, and to our pleasant surprise, we find that the nonvanishing photon angular momenta introduces certain ``dark" directions at the center of the image of the source in the observer's sky. These directions shape the optical appearance of the source, and imply that the collapsing cloud casts a growing shadow in the observer's sky. This shadow resembles the well-studied black hole shadow.

Bardeen~\cite{jB73} describes the black hole shadow in the following way:  ``As seen by the distant observer, the black hole will appear as a `black hole' in the middle of the larger bright source". Cunningham and Bardeen~\cite{cCjB73} and more recent works ~\cite{hFfMeA00,rT04,aBrN06,sWtW07,aV00,kHkM09} studied this shadow for a number of important illuminating sources. These works show that the black hole shadow is determined by tracing backwards null geodesics from the observer's frame back to the illuminating source.\footnote{Current day millimeter-wave very-long baseline interferometric arrays such as the Event Horizon Telescope~\cite{EHT} are on the verge of observing this shadow for Sagittarius A$^*$, the supermassive black hole in the center of our galaxy.} Because the black hole region absorbs incoming radiation, certain directions in the observer's sky will remain dark and the collection of all these dark directions determines the shadow that the black hole casts.\footnote{In this work we assume that the black hole or the naked singularity results from the collapse of a bounded system. If, for instance, a maximally extended Schwarzschild manifold is considered instead, the possibility that an observer in the vicinity of future null infinity could receive light originating from the white hole singularity ought to be taken into account and in this event the shadow would have different characteristics. We do not consider this possibility in this work based on the instability of white holes as demonstrated by Eardley~\cite{dE74}.},\footnote{We mention here that other ultracompact objects such as gravastars also generate shadows and in fact they may mask a black hole shadow. For recent work concerning this challenge, see, for instance, Refs.~\cite{cClR07,cClR08,nShStT14,hSaFcYyN15}.} Clearly, as long as the black hole is an equilibrium state, then its shadow remains steady provided the external illumination remains steady as well. For the case considered in this work, the changing background geometry adds a transient character to the perceived shadow. Nevertheless, the back-ray-tracing technique employed in the studies of black hole shadows is applicable to our problem as well. Through this technique, we compute the frequency shift suffered by photons and also study the characteristics of the shadow.  

Although the optical appearance and shadow associated with the formation of an event horizon have been studied a long time ago (see for instance Refs.~\cite{wAkT68,jJ69,kLrR79,hLvF06,vFkKhL07}), no particular attention has been placed on the effects that the formation of a naked singularity has on the eyes of an asymptotic observer.\footnote{A notable exception constitutes the work in Ref.~\cite{kNnKhI03}. We shall discuss further their approach and results in the conclusions section.} It is {\it a priori} not clear whether the formation of a naked singularity casts a shadow. If it does cast a shadow, one might wonder what its characteristic properties are. We delineate these issues by comparing the shadow that the collapsing cloud casts when the end state is a black hole to that alternative scenario, i.e. the case of a naked singularity. We find that in the former case the photosphere plays a crucial role as has been pointed out long ago by Ames and Thorne~\cite{wAkT68}. Photons around the photosphere are responsible for the optical appearance, while photons grazing the horizon are received by the asymptotic observer as exponentially redshifted. In contrast, when a  Cauchy horizon forms, the ``visible'' part of the singularity plays the dominant role in shaping up the shadow. The role that the singularity plays in the formation of the shadow is traced to a particular property of nakedly singular Tolman-Bondi spacetime. In such a spacetime, nonradial, future directed null geodesics escape from the central singularity and reach an asymptotic observer. This property has been proven for the case of marginally bounded Tolman-Bondi spacetime by Mena and Nolan~\cite{bNfM01,bNfM02}. Because of the important role that these null geodesics play in the formation of the shadow, we have recently verified their existence for the case of a nakedly self-similar Tolman-Bondi model~\cite{nOoStZ15b}, and in the present work we show that Mena and Nolan's conclusion remains valid for the large class of \emph{bounded} nakedly singular Tolman-Bondi models. For the case of a Cauchy horizon formation, and once the observer lies to the future of this horizon, by a combination of analytical and numerical techniques we separate the past directed null geodesics emanating from the observer into those reaching the singularity and those reaching the asymptotic source. At first, we find that photons emanating from the central singularity and reaching the observer are infinitely redshifted and thus the singularity is invisible to the eyes of the asymptotic observer. In addition, we estimate the redshift suffered by the important class of photons that originate from the site of the asymptotic source, traverse the collapsing cloud and eventually reach the observer. We compute this redshift as a function of the viewing angles as perceived in the frame of the asymptotic observer and thus we extend the analysis in~I where the considerations were restricted to radial photons. However, beyond the redshift analysis, the structure of the shadow as perceived by the asymptotic observer adds a new important component to the proposal put forward in~I. This component, with potentially observational consequences, is related to the diverse time scales required by the shadow to develop in the eyes of the asymptotic observer. By an analysis of the snapshots of the image as perceived by the observer, we find that while for the case of a black hole formation, a few crossing times are sufficient for the shadow to occult the source from the eyes of the observer, this is not the case once a Cauchy horizon forms. It takes considerably many more crossing times so that the source can eventually be occulted from the observer's eyes. Although for the case of stellar collapse the crossing time is extremely short, the situation is different for seeds leading to supermassive black holes. These issues are discussed in the conclusion section of the paper.
 
The structure for the remainder of this paper is the following. In Sec.~\ref{Sec:NullGeo}, because of the importance of null geodesics in this work, we begin with a brief derivation of the relevant equations of motion describing null geodesics (with and without angular momentum) on an arbitrary spherically symmetric background spacetime, and we derive the necessary formulas for the redshift of photons as measured by a certain class of preferred observers. Next, we specialize to the case of Tolman-Bondi spacetimes, describing the collapse of spherical dust clouds forming a locally or globally naked singularity. Based on methods from the theory of dynamical systems, we show the existence of infinitely many radial and nonradial null geodesics emanating from the singularity and analyze their local behavior. In particular, we prove that photons emitted from dust particles lying arbitrarily close to the central singularity suffer an infinite redshift. Next, in Sec.~\ref{Sec:ID}, we specify the initial data in our collapse model, and we provide a sufficient condition on the initial compactness ratio of the cloud under which the collapse leads to the formation of a naked singularity instead of a black hole. In Sec.~\ref{Sec:Raytracing} we present a detailed discussion for our ray tracing method which ultimately allows us to determine the redshift image perceived by an asymptotic static observer. One of the key steps here consists in the determination of the critical angle of incidence $\hat{\alpha}$, which separates the null geodesics emanating from the naked singularity from those originating from the asymptotic source. Finally, in Sec.~\ref{Sec:Results} we present the redshift images for a dust cloud forming either a black hole or a globally visible naked singularity, and compare the two cases. Conclusions are drawn in Sec.~\ref{Sec:Conclusions}, and the article ends with the inclusion of two appendixes. In Appendix~\ref{App:Proof} the proof of a technical result needed for establishing the existence of nonradial null geodesics emanating from the singularity is given. In Appendix~\ref{App:SS}, we discuss the aspects of the shadow of a naked singularity as seen by a comoving, free-falling observer in the particular case of self-similar Tolman-Bondi collapse, and we provide an analytic formula for the critical angle $\hat{\alpha}$.

\section{Null geodesics in spherically symmetric collapsing spacetimes}
\label{Sec:NullGeo}

We begin this section by reviewing basic properties of null geodesics propagating on an arbitrary spherically symmetric spacetime. In particular, we point out the existence of a preferred vector field ${\bf X}$, the Kodama vector field~\cite{hK80}, which reduces to the asymptotically unit timelike Killing vector field in the Schwarzschild case. We also show that the total redshift of photons measured by observers moving along integral curves of ${\bf X}$ assumes a particular simple form. Next, we specialize our considerations to the case of Tolman-Bondi dust collapse, and analyze the null geodesic flow in the vicinity of the naked singularity. Using standard results from the theory of dynamical systems, we show the existence of infinitely many radial and nonradial null geodesics emanating from the central singularity, a result that will play an important role in the considerations of Sec.~\ref{Sec:Raytracing}. Finally, we prove that photons emitted from dust particles lying arbitrarily close to the central singularity suffer an infinite redshift when they travel along such null geodesics and are eventually detected by an asymptotic observer.

\subsection{General framework}

Consider an arbitrary spherically symmetric spacetime metric of the form
\begin{equation}
{\bf g} = -e^{2\Phi(t,R)} dt^2 + e^{2\Psi(t,R)} dR^2 
 + r(t,R)^2\left( d\vartheta^2 + \sin^2\vartheta d\varphi^2 \right),
\label{Eq:Metric}
\end{equation}
where $\Phi$, $\Psi$ and $r$ are functions of the local coordinates $(t,R)$, and where $\vartheta$ and $\varphi$ are angular coordinates on the two-sphere. We are concerned with null geodesics propagating on the spacetime described by the metric given in Eq.~(\ref{Eq:Metric}). Because of rotational invariance, we can assume without loss of generality that the null geodesics are confined to the equatorial plane $\vartheta = \pi/2$. The Lagrangian describing the motion is given by
\begin{equation}
{\cal L}\left( t,R,\varphi,
\frac{dt}{d\lambda},\frac{dR}{d\lambda},\frac{d\varphi}{d\lambda}ÊÊ\right)
 = \frac{1}{2}\left[ -e^{2\Phi(t,R)}\left( \frac{dt}{d\lambda} \right)^2
 + e^{2\Psi(t,R)}\left( \frac{dR}{d\lambda} \right)^2
 + r(t,R)^2\left( \frac{d\varphi}{d\lambda} \right)^2 \right],
\end{equation}
where $\lambda$ is an affine parameter. Since $\varphi$ is cyclic, we have conservation of total angular momentum,
\begin{equation}
\ell := \frac{\partial\cal L}{\partial\left( \frac{d\varphi}{d\lambda} \right)} 
 = r^2 \frac{d\varphi}{d\lambda},
\end{equation}
and the Euler-Lagrange equations yield the first-order system
\begin{eqnarray}
\frac{dt}{d\lambda} &=& e^{-2\Phi}\pi^t,\\
\frac{d\pi^t}{d\lambda} &=& \dot{\Phi} e^{-2\Phi}(\pi^t)^2 - \dot{\Psi} e^{-2\Psi}(\pi^R)^2
 - \frac{\ell^2}{r^3}\dot{r},\\
\frac{dR}{d\lambda} &=& e^{-2\Psi}\pi^R,\\
\frac{d\pi^R}{d\lambda} &=& -\Phi' e^{-2\Phi}(\pi^t)^2 + \Psi' e^{-2\Psi}(\pi^R)^2
 + \frac{\ell^2}{r^3} r', 
\end{eqnarray}
where we have introduced the momenta $\pi^t := e^{2\Phi} dt/d\lambda$ and $\pi^R := e^{2\Psi} dR/d\lambda$ and where the dot and the prime denote partial differentiation with respect to $t$ and $R$, respectively. These equations should be supplemented with the nullity constraint ${\cal L} = 0$.

For the following, we introduce the vector field
\begin{equation}
{\bf X} := e^{-(\Phi + \Psi)}
\left( r'\frac{\partial}{\partial t} - \dot{r}\frac{\partial}{\partial R} \right)
\end{equation}
and the Misner-Sharp mass function~\cite{cMdS64} $m$ defined by
\begin{equation}
1 - \frac{2m}{r} = {\bf g}(dr,dr) = -e^{-2\Phi}\dot{r}^2 + e^{-2\Psi} r'^2.
\label{Eq:MisnerSharp}
\end{equation}
Up to a sign, the vector field ${\bf X}$ is uniquely characterized by the properties of being orthogonal to the invariant two-spheres, of leaving the surfaces of constant areal radius invariant, i.e. ${\bf X}[r] = 0$, and of being normalized such that
\begin{equation}
{\bf g}({\bf X},{\bf X}) = -\left( 1 - \frac{2m}{r} \right).
\label{Eq:Xnorm}
\end{equation}
The existence of this natural vector field has been noticed long ago by Kodama~\cite{hK80} in a different context. It follows that ${\bf X}$ is timelike outside the apparent horizon $r > 2m$. Associated with this vector field is the energy quantity
\begin{equation}
{\cal E} := -{\bf g}({\bf X},{\bf p}) = e^{-(\Phi + \Psi)}\left(r'\pi^t + \dot{r}\pi^R \right),
\label{Eq:DefE}
\end{equation}
where here
\begin{equation}
{\bf p} = \frac{dt}{d\lambda}\frac{\partial}{\partial t}
 + \frac{dR}{d\lambda}\frac{\partial}{\partial R}
 + \frac{d\varphi}{d\lambda}\frac{\partial}{\partial \varphi}
 = e^{-2\Phi}\pi^t\frac{\partial}{\partial t}
 + e^{-2\Psi}\pi^R\frac{\partial}{\partial R}
 + \frac{\ell}{r^2}\frac{\partial}{\partial \varphi}
\label{Eq:Defp}
\end{equation}
denotes the four-momentum of the particle. An additional useful equation is the change of areal radius $r$ along any null geodesics, given by
\begin{equation}
\frac{dr}{d\lambda} = \dot{r}e^{-2\Phi}\pi^t + r' e^{-2\Psi}\pi^R.
\label{Eq:drdlambda}
\end{equation}
Combining Eqs.~(\ref{Eq:MisnerSharp}),~(\ref{Eq:DefE},~and~(\ref{Eq:drdlambda}) and using ${\cal L} = 0$ yields the equation
\begin{equation}
\left( \frac{dr}{d\lambda} \right)^2 
 + \frac{\ell^2}{r^2}\left( 1 - \frac{2m}{r} \right) = {\cal E}^2.
\label{Eq:ECons}
\end{equation}
In the vacuum case, $m$ is constant and ${\bf X}$ reduces to the time-translation Killing vector field. In this case, the energy quantity ${\cal E}$ is constant along the null geodesic flow and Eq.~(\ref{Eq:ECons}) expresses the familiar effective equation describing the radial motion for null geodesics.

Finally, for later use, we compute the frequency shift of photons with respect to observers moving along integral curves of ${\bf X}$. This frequency shift is given by
$$
\frac{\nu_{obs}}{\nu_e} = \frac{{\bf g}({\bf U},{\bf p})_{obs}}{{\bf g}({\bf U},{\bf p})_e},
\qquad {\bf U} = \frac{{\bf X}}{\sqrt{-{\bf g}({\bf X},{\bf X})}},
$$
which is valid as long as ${\bf X}_{obs}$ and ${\bf X}_e$ are timelike, such that both the observer's and the emitter's four-velocity ${\bf U}$ is well defined. Using Eqs.~(\ref{Eq:Xnorm}) and (\ref{Eq:DefE}) we obtain
\begin{equation}
\frac{\nu_{obs}}{\nu_e} = \sqrt{
 \frac{\left. 1 - \frac{2m}{r} \right|_e }{\left. 1 - \frac{2m}{r} \right|_{obs} }}
\frac{{\cal E}_{obs}}{{\cal E}_e}.
\label{Eq:XRedshift}
\end{equation}
In the Schwarzschild case, ${\cal E}$ is constant along the null rays, and we recover the familiar expression describing the gravitational redshift. For a photon traveling from past to future null infinity through a collapsing spherical object of finite radius, Eq.~(\ref{Eq:XRedshift}) yields the remarkably simple formula
\begin{equation}
\frac{\nu_\infty^+}{\nu_\infty^-} = \frac{{\cal E}_1^+}{{\cal E}_1^-},
\label{Eq:Total_redshift}
\end{equation}
where ${\cal E}_1^+$ (${\cal E}_1^-$) is the energy ${\cal E}$ of the photons at the moment they exit (enter) the collapsing object and $\nu_\infty^+$ and $\nu_\infty^-$ refer to the frequencies at future and past null infinity, respectively, as measured by asymptotic static observers. We will use this formula in the next section.

\subsection{Null geodesics in the Tolman-Bondi spacetime}

For the particular case of dust collapse, it is convenient to choose $(t,R)$ as synchronous comoving coordinates~\cite{MTW-Book} adapted to the flow of the dust particles. For such coordinates we have in Eq.~(\ref{Eq:Metric}), $\Phi = 0$, $t = \tau$ is the proper time along the trajectories of the dust particles, and $e^{\Psi(\tau,R)} = 1/\gamma(\tau,R) = r'(\tau,R)/\sqrt{1 + 2E(R)}$. The areal radius $r(\tau,R)$ of the spherical dust shell labeled by $R$ at proper time $\tau$ is determined by the equation
\begin{equation}
\frac{1}{2}\dot{r}^2 - \frac{m(R)}{r} = E(R),
\label{Eq:FreeFall}
\end{equation}
with $E(R)$ the energy of the dust shell $R$, see Ref.~\cite{nOoS11} for notation and further properties of the functions $r(\tau,R)$, $\dot{r}$, $r'$ and $\gamma$. Hereafter, we require the conditions~(i)--(viii) in Ref.~\cite{nOoS11} to hold.  These requirements imply regularity of the initial data and the absence of shell-crossing singularities (among other reasonable physical properties), and they lead to the formation of a locally naked shell-focusing singularity.

For the following, we use the nullity constraint ${\cal L} = 0$ in order to replace $(\pi^\tau,\pi^R)$ with the single function
\begin{equation}
\Pi := \gamma\frac{\pi^R}{\pi^\tau},
\end{equation}
in terms of which
\begin{equation}
\pi^\tau = \frac{|\ell|}{r}\frac{1}{\sqrt{1 - \Pi^2}},\qquad
\pi^R = \frac{1}{\gamma}\frac{|\ell|}{r}\frac{\Pi}{\sqrt{1 - \Pi^2}},
\label{Eq:pitauR}
\end{equation}
where we have taken the absolute value of $\ell$ in order to make sure that $\pi^\tau > 0$ such that the trajectory is future directed.\footnote{Note that in the radial case $\Pi = \pm 1$ and $\ell=0$. In this case, the equations describing the null geodesics are simply given by $d\tau = \Pi^{-1} dR/\gamma$.} In terms of this new variable, the equations describing the null geodesics are: 
\begin{eqnarray}
\frac{d\tau}{d\lambda} &=& \frac{|\ell|}{r}\frac{1}{\sqrt{1 - \Pi^2}},\\
\frac{dR}{d\lambda} &=& 
 \frac{|\ell|}{r}\frac{\sqrt{1+2E}}{r'}\frac{\Pi}{\sqrt{1 - \Pi^2}},\\
\frac{d\Pi}{d\lambda} &=& \frac{|\ell|}{r}\sqrt{1 - \Pi^2}
\left[ \frac{1}{r}\sqrt{1 + 2E} +
 \left( \frac{\dot{r}}{r} - \frac{\dot{r}'}{r'} \right)\Pi \right],
\end{eqnarray}
and the change of the areal radius $r$ along the null geodesic is described by
\begin{equation}
\frac{dr}{d\lambda} = \frac{|\ell|}{r}\frac{1}{\sqrt{1 - \Pi^2}}
\left( \dot{r} + \sqrt{1+2E}\,\Pi \right).
\end{equation}
For the following analysis, we reparametrize the geodesics in terms of the coordinate $R$. This change of parametrization is legitimate as long as $\Pi\neq 0$. Under this change of parametrization, the above equations take the form
\begin{eqnarray}
\frac{d\tau}{dR} &=& \frac{r'}{\sqrt{1+2E}}\frac{1}{\Pi},\\
\frac{d\varphi}{dR} &=& \frac{\sign(\ell)}{r}\frac{r'}{\sqrt{1+2E}}\frac{\sqrt{1-\Pi^2}}{\Pi},\\
\frac{dr}{dR} &=& r'\left( 1 + \frac{\dot{r}}{\sqrt{1+2E}}\frac{1}{\Pi} \right),
\label{Eq:drdR}\\
\frac{d\Pi}{dR} &=& (1-\Pi^2) r'\left[
 \frac{1}{\sqrt{1+2E}} \left( \frac{\dot{r}}{r} - \frac{\dot{r}'}{r'} \right) + \frac{1}{r}\frac{1}{\Pi}
 \right].
\label{Eq:dPidR}
\end{eqnarray}
Note that these equations are also valid in the case of radial null geodesics, for which $\Pi = \pm 1$. The last two equations form a closed system of nonlinear ordinary differential equations for the variables $(r,\Pi)$, since the coefficients $E,r,\dot{r},r',\dot{r}'$ appearing on the right-hand side of Eqs~(\ref{Eq:drdR})~and~(\ref{Eq:dPidR}) can be expressed in terms of $R$ and $r$ only.

\subsection{Radial and nonradial null geodesics emanating from the singularity}
\label{Sec:rays_from_sing}

In order to analyze the behavior of the solutions in the vicinity of the central singularity $(\tau,R) = (0,0)$, it turns out to be convenient to replace the areal radius $r$ with the new variable
$$
\chi := \sqrt{\frac{2m(R)}{r}},
$$
and the coordinate radius $R$ labeling the dust shells with $s := 1/R$. Note that $\chi = 1$ characterizes the location of the apparent horizon, while $0\leq \chi < 1$ defines the region lying to the past of it which includes the central singularity. Equations~(\ref{Eq:drdR})~and~(\ref{Eq:dPidR}) can be rewritten as the following nonautonomous dynamical system for the variables $(\chi,\Pi)$: 
\begin{equation}
\frac{d}{ds}\left( \begin{array}{c} \chi \\ \Pi \end{array} \right)
 = \left. X(R,\chi,\Pi) \right|_{R = 1/s}
 = \left. \left( \begin{array}{c} X_1(R,\chi,\Pi) \\ X_2(R,\chi,\Pi) \end{array} \right)
 \right|_{R = 1/s},
\label{Eq:DynSys}
\end{equation}
with the functions $X_1$ and $X_2$ given by
\begin{eqnarray*}
X_1(R,\chi,\Pi) &:=& \frac{\chi}{2}\left\{ 
\left[ R + \frac{\chi^3}{c(R)^{3/2}}\sqrt{1 + 2E(R)\chi^{-2}}
  \Lambda\left(\frac{R\sqrt{c(R)}}{\chi},R \right) \right]
   \left[ 1 - \sqrt{\frac{1 + 2E(R)\chi^{-2}}{1 + 2E(R)}}\frac{\chi}{\Pi} \right]
 - 3R - R^2\frac{c'(R)}{c(R)} \right\},\\
X_2(R,\chi,\Pi) &:=& (1-\Pi^2)\\
 &\times& \left\{ 
\frac{\chi^3}{c(R)^{3/2}}\Lambda\left(\frac{R\sqrt{c(R)}}{\chi},R \right)
 \left[ \frac{3 + 4E(R)\chi^{-2}}{2\sqrt{1 + 2E(R)}}\chi
 - \sqrt{1 + 2E(R)\chi^{-2}} \frac{1}{\Pi} \right]
 - R K\left(\frac{R\sqrt{c(R)}}{\chi},R \right)\chi - \frac{R}{\Pi}  \right\},
\end{eqnarray*}
where the functions $c(R)$, $\Lambda(y,R)$ and $K(y,R)$ are defined by
\begin{eqnarray*}
c(R) &:=& \frac{2m(R)}{R^3},\\
q(R) &:=& \sqrt{-\frac{R E(R)}{m(R)}},\\
K(y,R) &:=& \frac{1}{\sqrt{1 + 2E(R)}}\left[ \sqrt{1 - q(R)^2 y^2} \frac{R c'(R)}{2c(R)}
 - \frac{q(R)^2 y^2}{\sqrt{1 - q(R)^2 y^2}} \frac{R q'(R)}{q(R)} \right],\\
\Lambda(y,R) &:=& 2\frac{q'(R)}{Rq(R)} h(q(R),y) 
  - \frac{c'(R)}{2Rc(R)} g(q(R),y),\\
h(q,y) &:=& \frac{1}{\sqrt{1-q^2}} - \frac{y^3}{\sqrt{1- q^2y^2}} - \frac{3}{2}g(q,y),\\
g(q,y) &:=& \frac{ f(qy) - f(q) }{q^3},\\
f(x) &:=& x\sqrt{1 - x^2} + \arccos(x),
\end{eqnarray*}
with $y := \sqrt{r/R}$, see Ref.~\cite{nOoS11} for more details and properties of these functions.

\begin{figure}[h!]
\begin{center}
\includegraphics[width=8.5cm]{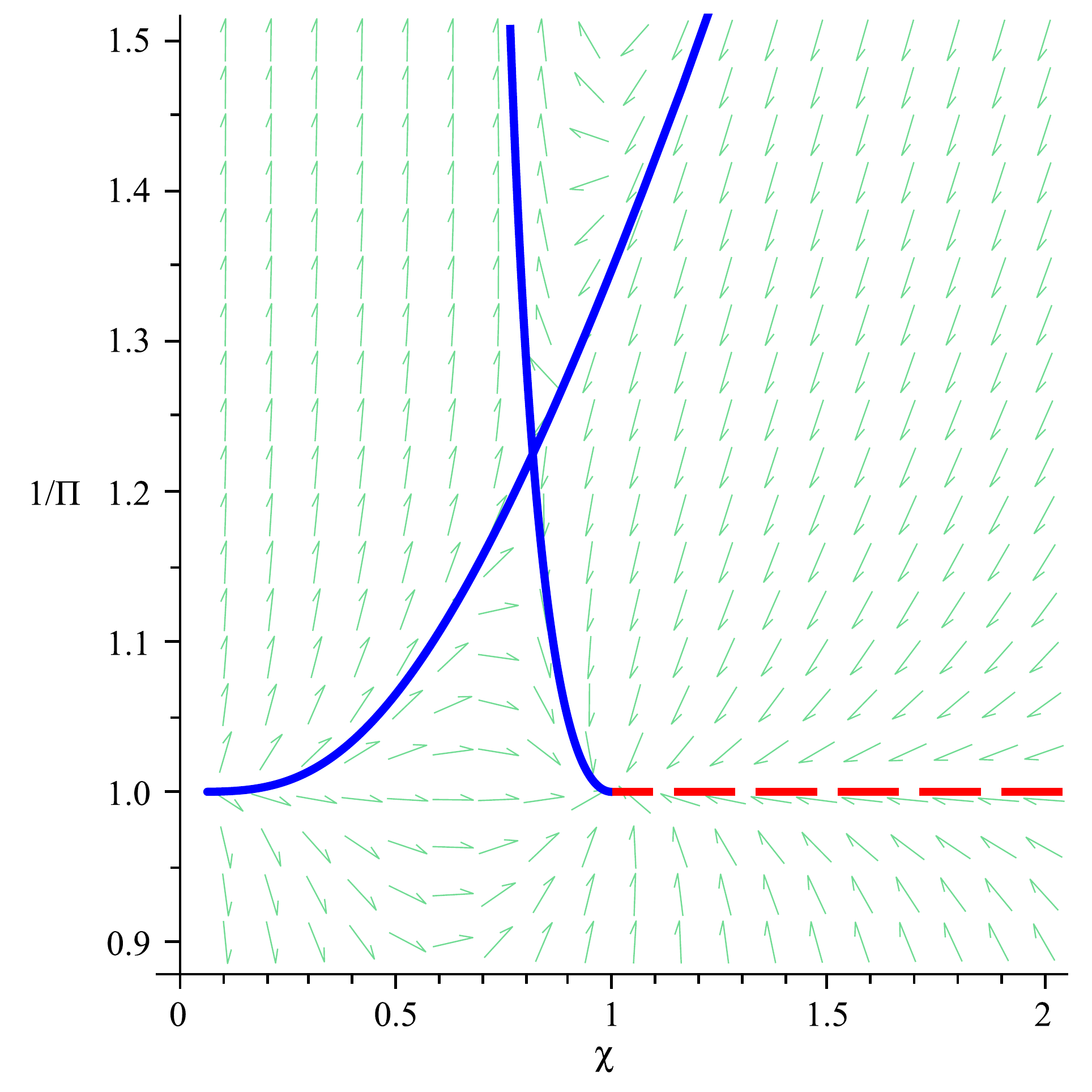}
\end{center}
\caption{\label{Fig:PhasePortrait} The phase portrait for the autonomous dynamical system defined in Eq.~(\ref{Eq:DynSysInf}). Note that we use the variable $1/\Pi$ instead of $\Pi$ in order to have a better view of the phase space in the region $0 < \Pi \leq 1$. The arrows show the direction of the vector field, the blue solid lines are the stable and unstable local manifolds associated with the critical point $(\chi_2,\Pi_2) = (\sqrt{2/3},\sqrt{2/3})$ and the red dashed line is an integral curve corresponding to a radial light ray. As the figure suggests, the basin of attraction for the critical point $(\chi_1,\Pi_1) = (1,1)$ is the region that lies below the stable manifold.}
\end{figure}

We are particularly interested in the behavior of the solutions of Eq.~(\ref{Eq:DynSys}) in the vicinity of the central singularity $(\tau,R) = (0,0)$. Notice that in terms of the above notation,
$$
\chi = R\sqrt{c(R)}\sqrt{\frac{R}{r}},
$$
such that a finite, positive limit of $\chi$ as $R\to 0$ implies $r/R\to 0$. Therefore, any solution of Eq.~(\ref{Eq:DynSys}) for which $\chi$ converges to a finite, positive value when $R\to 0$ corresponds to a null geodesics emanating or terminating at the central singularity. In order to understand the limit $R\to 0$ (or $s\to \infty$) we first analyze the autonomous system which is obtained by setting $R=0$ in the right-hand side of Eq.~(\ref{Eq:DynSys}), that is,
\begin{equation}
\frac{d}{ds}\left( \begin{array}{c} \chi \\ \Pi \end{array} \right)
 = X(0,\chi,\Pi)
 = A_0\chi^3\left( \begin{array}{r} 
 \chi\left( 1 - \frac{\chi}{\Pi} \right)Ê\\
 (1-\Pi^2)\left( 3\chi - \frac{2}{\Pi} \right) 
\end{array} \right),
\label{Eq:DynSysInf}
\end{equation}
where we have used the assumption $E(0) = 0$ which follows from condition (i) in Ref.~\cite{nOoS11}, and where $A_0 := \Lambda(0,0)/(2c(0)^{3/2}) > 0$. There are two critical points of this system with $\chi > 0$, namely
$$
(\chi_1,\Pi_1) = (1,1),\qquad
(\chi_2,\Pi_2) = \sqrt{\frac{2}{3}}(1,1).
$$
The linearization about these points (keeping $R=0$ fixed) is
$$
DX(0,\chi_1,\Pi_1) = A_0\left( \begin{array}{rr}
-1 & 1 \\ 0 & -2
\end{array} \right),\qquad
DX(0,\chi_2,\Pi_2) = A_0\left( \frac{2}{3} \right)^{3/2}
\left( \begin{array}{rr}
-1 & 1 \\ 1 & 1
\end{array} \right).
$$
In the first case, both eigenvalues are negative and thus the critical point $(\chi_1,\Pi_1)$ is an attractor for the system~(\ref{Eq:DynSysInf}). In the second case, the eigenvalues are $\pm\sqrt{2}(2/3)^{3/2} A_0$, and hence the critical point $(\chi_2,\Pi_2)$ is hyperbolic with associated one-dimensional stable and unstable local manifolds. A phase portrait for the system~(\ref{Eq:DynSysInf}) is shown in Fig.~\ref{Fig:PhasePortrait}.

After understanding the behavior of the autonomous system, we return to the nonautonomous system defined in Eq.~(\ref{Eq:DynSys}), which we rewrite as
\begin{equation}
\frac{d}{ds}\left( \begin{array}{c} \chi \\ \Pi \end{array} \right)
 = X(0,\chi,\Pi) + \left. R Y(R,\chi,\Pi) \right|_{R = 1/s},
\end{equation}
where
$$
Y(R,\chi,\Pi) := \frac{1}{R}\left[ X(R,\chi,\Pi) - X(0,\chi,\Pi) \right],\qquad R > 0.
$$
It is not difficult to verify that for small enough $\varepsilon > 0$ and a suitable open neighborhood $U$ of $(\chi,\Pi) = (1,1)$ in $\Real^2$, the map $X: (0,\varepsilon)\times U\to \Real^2, (R,\chi,\Pi)\mapsto X(R,\chi,\Pi)$ defined in Eq.~(\ref{Eq:DynSys}) is continuous. Consequently, $Y: (0,\varepsilon)\times U\to \Real^2, (R,\chi,\Pi)\mapsto Y(R,\chi,\Pi)$ defines a continuous map, and it is not difficult to prove that this map is bounded. Now we resort to the following general result whose proof we give in Appendix~\ref{App:Proof} for completeness.

\begin{theorem}[Stability theorem, cf. Theorem 4.5 in Ref.~\cite{BrauerNohel-Book}]
\label{Thm:Stability}
Let $U\subset \Real^m$ be an open subset and let $X: U\to \Real^m, x\mapsto X(x)$ be a continuously differentiable map and $x^*\in U$ a point such that
$$
X(x^*) = 0
$$
and such that all the eigenvalues of the linearization $DX(x^*): \Real^m\to \Real^m$ at $x^*$ have a negative real part. Let $Y: (0,\infty)\times U\to \Real^m, (s,x)\mapsto Y(s,x)$ be a bounded continuous map.

Then, there exists $s_1 > 0$ and an open subset $V\subset U$ containing $x^*$ such that any (maximally extended) solution $x(s)$ of the Cauchy problem
\begin{equation}
\left\{ \begin{array}{l}
\frac{dx}{ds} = X(x) + \frac{1}{s} Y(s,x),\quad s\geq s_0,\\
x(s_0) = x_0
\end{array} \right.
\label{Eq:PertDyn}
\end{equation}
with $(s_0,x_0)\in (s_1,\infty)\times V$ exists for all $s\geq s_0$ and satisfies
$$
\lim\limits_{s\to\infty} x(s) = x^*.
$$
\end{theorem}

{\bf Remark:} $x(s) = x^*$ for all $s\geq 0$ is not necessarily a solution of the Cauchy problem~(\ref{Eq:PertDyn}) with $x_0 = x^*$, since we do not assume that $Y(s,x^*) = 0$.

According to Theorem~\ref{Thm:Stability}, there exists an open neighborhood $V$ of the point $(\chi,\Pi) = (1,1)$ and a value $R_0 > 0$, such that any solution $(\chi(s),\Pi(s))$ of the nonautonomous system~(\ref{Eq:DynSys}) with $\left. (\chi(s),\Pi(s)) \right|_{s=1/R_0}\in V$ converges to the point $(1,1)$ as $s\to \infty$. Physically, this implies that any null geodesic passing close enough to the point $(R,\chi,\Pi) = (0,1,1)$ emanates from the central singularity. In particular, we have shown that there exists infinitely many null geodesics emanating from the central singularity, with or without angular momentum. As it will turn out, this property will be of central importance to the main results of this paper and its implications will become clear in Sec.~\ref{Sec:Raytracing}.

Clearly, the result described in the previous paragraph does not provide a full picture for the behavior of null geodesics in a vicinity of the naked singularity, since it leaves open the possibility for the existence of null geodesics emanating from the singularity with asymptotic values for $(\chi,\Pi)$ different from $(1,1)$. A full picture in the generic Tolman-Bondi collapse case analyzed here is beyond the scope of the paper. For a full qualitative analysis of the null geodesic flow in the particular case of a nakedly singular self-similar Tolman-Bondi spacetime we refer the reader to our recent work~\cite{nOoStZ15b}; see also Appendix~\ref{App:SS}.

\subsection{Infinite redshift}
\label{Sec:Infinite_redshift}

After having showed the existence of infinitely many null geodesics, with and without angular momentum, emanating from the central singularity, we analyze the frequency shift the photons undergo along these geodesics as measured by free-falling observers. In particular, we show that photons emitted from a region very close to the central singularity are infinitely redshifted.

The frequency shift along outgoing radial null geodesics is given by~\cite{nOoS14}
\begin{equation}
\frac{\nu_{obs}}{\nu_e} = \exp
\left[ \int\limits_{R_e}^{R_{obs}} \frac{\dot{\gamma}}{\gamma^2}(\tau_+(R),R) dR \right],
\label{Eq:Redshift}
\end{equation}
where $R\mapsto (\tau_+(R),R)$ parametrizes the outgoing radial light ray, and $R = R_e$ and $R = R_{obs}$ denote the location of the free-falling emitter and observer, respectively. It follows from the explicit expressions for $\dot{\gamma}/\gamma^2$ derived in Lemma~3 of~\cite{nOoS14} that $\dot{\gamma}/\gamma^2 \simeq -1/R^2$ for small $R$, along radial light rays emanating from the central singularity with $\chi\to 1$ as $R\to 0$, and consequently, there is an infinite redshift as $R_e\to 0$, in agreement with Ref.~\cite{dC84}.

Regarding the frequency shift along nonradial null geodesics emanating from the central singularity such that $(\chi,\Pi)\to (1,1)$ as $R\to 0$, we have
\begin{equation}
\frac{\nu_{obs}}{\nu_e} = \frac{{\bf g}({\bf u},{\bf p})_{obs}}{{\bf g}({\bf u},{\bf p})_e}
 = \frac{r_e}{r_{obs}}\sqrt{ \frac{1 - \Pi_e^2}{1 - \Pi_{obs}^2} },
\end{equation}
where we have used Eqs.~(\ref{Eq:Defp}) and~(\ref{Eq:pitauR}). Note that for a free-falling observer at $R = R_{obs} > 0$, we have $r_{obs} > 0$ and $\Pi_{obs}^2\neq 1$, since the null ray is nonradial. However, as we consider an emitter on a free-falling trajectory located closer and closer to the central singularity, $r_e\to 0$ and $\Pi_e\to 1$, and we obtain again an infinite redshift.

\section{Initial conditions for the collapsing cloud}
\label{Sec:ID}

In this section, we briefly discuss the family of initial data used in our analysis. The initial data for the collapsing dust cloud consist in the initial density profile $\rho(R)$ and the initial profile for the radial velocity $v(R)$ (for a precise definition of this velocity see~\cite{nOoS11}). For simplicity, we choose time-symmetric initial data for the dust collapse, implying that $v(R) = 0$ for all $R\geq 0$. To be in accord with the simulations presented in~I (Ref.~\cite{nOoStZ14}), the density profile is chosen in the following form:
\begin{equation}
\rho(R) = \rho_c\times\left\{ \begin{array}{ll} 
 1 - \frac{5}{3}a_0\left( \frac{R}{R_1} \right)^2 + \frac{7}{3}a_1 \left( \frac{R}{R_1} \right)^4 + 3a_2 \left( \frac{R}{R_1} \right)^6 + \frac{11}{3}a_3\left( \frac{R}{R_1} \right)^8, & 0 \leq R \leq R_1,\\
0,& R > R_1, \end{array} \right.
\label{Eq:Density_profile}
\end{equation}
where $R_1 > 0$ is the initial areal radius of the cloud, $\rho_c > 0$ is the central density, and $a_0$ is a dimensionless parameter restricted to the interval $0 < a_0 \leq 1$ which characterizes the flatness of the initial density $\rho$. The other parameters are given in terms of $a_0$ by $a_1 = -3(6 - 5a_0)/7$, $a_2 = (8-5a_0)/3$, and $a_3 = -(9 - 5a_0)/11$. The initial density profile defined in Eq.~(\ref{Eq:Density_profile}) is continuous and monotonously decreasing on the interval $[0,\infty)$, and provided $0 <  a_0 < 12/5$, it satisfies the assumptions (i)--(viii) listed in Sec.~II of Ref.~\cite{nOoS11} such as smoothness, boundedness, non-negative mass density, absence of shell-crossing singularities, absence of initially trapped surfaces, with the exception of $C^\infty$-differentiability at the surface of the cloud.\footnote{Nevertheless, the density profile is twice continuously differentiable everywhere.} In particular, the profile given in Eq.~(\ref{Eq:Density_profile}) is smooth at the center where indeed $\rho'(0) = 0$ and $\rho''(0) < 0$. As shown by Christodoulou~\cite{dC84}, these conditions lead to the formation of a shell-focusing singularity which is visible at least to local observers.

The initial compactness ratio ${\cal C} := 2m_1/R_1$ of a collapsing cloud with initial density given by Eq.~(\ref{Eq:Density_profile}) turns out to be
\begin{equation}\label{Eq:m_1}
{\cal C} = \frac{8}{3}\pi R_1^2\rho_c \left( 1 - a_0 + a_1 + a_2 + a_3 \right).
\end{equation}
In our numerical simulations we fix $a_0 = 1$; nonetheless other values in the interval $(0,12/5)$ lead to the same qualitative results. Below, we make the particular choice ${\cal C} = 16/77 =: {\cal C}_0$ which leads to the formation of a black hole, and further consider initial data with ${\cal C} \leq {\cal C}_0/2$ which guarantee the formation of a globally  naked singularity; see paper~I.

\section{Back-ray tracing the photons from the observer}
\label{Sec:Raytracing}

\begin{figure}[h!]
\begin{center}
\includegraphics[width=9cm]{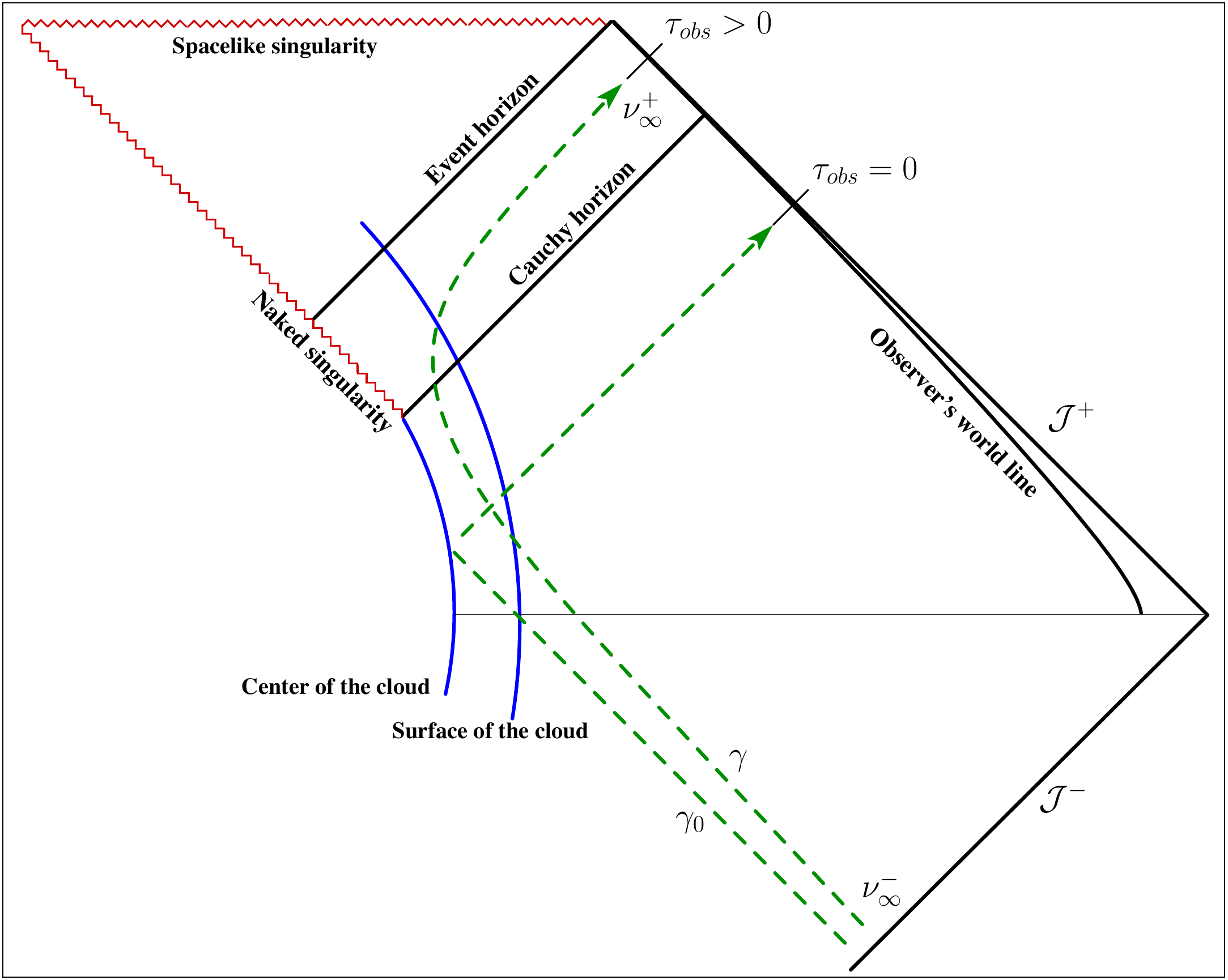}
\end{center}
\caption{\label{Fig:Conformal} Conformal diagram for the spherical collapse of a dust cloud forming a naked singularity. We illustrate the world line of the distant observer and the spacetime trajectories of two light rays $\gamma_0$ and $\gamma$. The particular radial light ray $\gamma_0$ is the one originating from the distant source which penetrates the cloud at the moment of time symmetry and reaches the observer at time $\tau_{obs} = 0$. $\gamma$ is an arbitrary trajectory of a null geodesic with angular momentum.}
\end{figure}

In the spacetime generated by the data described in the previous section, we consider a particular static observer at some distance $r_{obs} \gg 2m_1$ from the collapsing cloud who registers photons that have traversed the collapsing cloud. As in~I, we assume that the received radiation is generated by external sources distributed uniformly in the asymptotic region. This radiation does not interact with any intervening matter nor with the collapsing matter, and for simplicity we ignore any radiation generated within the collapsing cloud. However, unlike in~I, here we allow this radiation to have nonvanishing angular momentum. Thus, within the geometric optics approximation, photons composing this radiation are moving on null geodesics originating from the asymptotic source, going through the collapsing cloud and finally received by the detectors carried by our asymptotic observer. In the spirit of the proposal in~I, our focus in the next sections is to estimate the total frequency shift $\nu_\infty^+/\nu_\infty^-$ of the photons perceived by the observer, where hereafter $\nu_\infty^-$ and $\nu_\infty^+$ stand for the frequency of emission and detection, respectively.
In order to compute the frequency shift $\nu_\infty^+/\nu_\infty^-$ of these photons, we trace back their path along past-directed null geodesics intersecting the cloud. Each null geodesic is characterized by its impact parameter $b$ and the proper time $\tau_{obs}$ at which the observer measures its corresponding redshift. In Fig.~\ref{Fig:Conformal}, we illustrate a possible nonradial photon trajectory ($\gamma$) in the case where the collapse forms a naked singularity, and we also show the particular radial light ray ($\gamma_0$) penetrating the cloud at the moment of time symmetry, which we use as a reference to define $\tau_{obs} = 0$.

We divide the process of tracing back the null geodesics in three steps. The first step consists in tracing back the null rays from the observer to the surface of the cloud, and thus determining the areal radius $r_1^+$ of the surface of the cloud at the moment the ray penetrates it and its angle of incidence $\alpha$ (defined below). No numerical ray tracing is required in this step since we are dealing with null geodesics in a Schwarzschild spacetime, whose equations of motion are integrable.

The second step involves tracing back the light ray numerically from the surface of the cloud through the cloud itself. In the case where the object collapses to a black hole, the light ray reaches the surface of the cloud again in the past. However, in the naked singularity case, there is an alternative possibility that may occur, once the observer passes the Cauchy horizon. As we have shown in the previous section, there exist future directed null geodesics emanating from the naked singularity. Therefore, it is possible that the ray-traced null geodesic hits the naked singularity instead of extending to the distant source. A consequence of this property is that once the asymptotic observer passes the Cauchy horizon, he or she registers a smaller number of photons originating from the source. In this step, we determine numerically the critical angle $\hat{\alpha}$ which distinguishes between null geodesics emanating from the naked singularity from those that arrive from the external source at past null infinity.

Finally, in a third step, we vary the parameters $\tau_{obs}$ and $b$ characterizing the null geodesic. For each value of these parameters, we determine whether the null geodesic originates from the source or from the naked singularity. In the first case, we compute the total redshift the photons undergo, while in the latter case, as we showed in Sec.~\ref{Sec:Infinite_redshift}, the photons are infinitely redshifted.

In the following, we describe the details involved in realizing these three steps.

\subsection{Step 1: Tracing back the null rays from the observer to the surface of the cloud}
\label{Sec:Step1}

In the exterior region the null geodesics are described by Eq.~(\ref{Eq:ECons}) with constant energy ${\cal  E}$. Introducing the impact parameter $b := \ell/{\cal E}$ and rescaling the affine parameter such that $\ell^2 = 1$, Eq.~(\ref{Eq:ECons}) assumes the form
\begin{equation}
\left( \frac{dr}{d\lambda} \right)^2 + \frac{1}{r^2}\left( 1 - \frac{2m_1}{r} \right)
 = \frac{1}{b^2},
\label{Eq:EConsBis}
\end{equation}
where $m_1$ is the total mass of the cloud. This is the equation for a one-dimensional mechanical particle in the potential $(1 - 2m_1/r)/r^2$, whose maximum $1/(27m_1^2)$ is located at $r = 3m_1$ (the photosphere). In the following, we restrict ourselves to absolute values for the impact parameter $|b| < \overline{b}$, lying below the critical value 
$$
\overline{b} = \sqrt{27}m_1,
$$
corresponding to the maximum of the potential. Therefore, when $|b| < \overline{b}$, $1/b^2$ is larger than the maximum of the effective potential, implying that the null geodesic must intersect the cloud in its past, since otherwise it must have emanated from $r = 0$ in the Schwarzschild region which is not possible for an observer lying to the future of the reference curve $\gamma_0$, see Fig.~\ref{Fig:Conformal}.

Let $r = r_1^+$ denote the areal radius of the surface of the cloud at the event $x$ where the null geodesic intersects it. We define the angle $\alpha$ of incidence at $x$ in the following way: denote by
$$
{\bf e}_0 = \frac{\partial}{\partial\tau},\quad
{\bf e}_1 = \gamma\frac{\partial}{\partial R},\quad
{\bf e}_2 = \frac{1}{r}\frac{\partial}{\partial\vartheta},\quad
{\bf e}_3 = \frac{1}{r\sin\vartheta}\frac{\partial}{\partial\varphi}
$$
an orthonormal frame at $x$ adapted to a free-falling observer comoving with a dust particle at the surface of the cloud and chosen so that ${\bf e}_1$ is parallel to the outward pointing radial vector field at the event $x$.
Then, we define the incidence angle $\alpha$ by
\begin{equation}
{\bf p}_x = A\left( {\bf e}_0 - \cos\alpha\,{\bf e}_1 + \sin\alpha\, {\bf e}_3 \right),
\label{Eq:Defalpha}
\end{equation}
where $A > 0$ is a positive constant and ${\bf p}_{x}$ denotes the four-momentum of a photon which is restricted to move on the equatorial plane $\vartheta = \pi/2$. Geometrically, $\alpha$ determines the angle that the orthogonal projection of ${\bf p}_x$ onto the invariant plane  spanned by $({\bf e}_1,{\bf e}_3)$ makes with the radial outgoing direction ${\bf e}_1$. According to this parametrization, ${\bf p}_x$ is tangent to the ingoing (outgoing) radial geodesics for the particular value $\alpha = 0$ ($\alpha = \pi$).

Comparing Eq.~(\ref{Eq:Defalpha}) with Eq.~(\ref{Eq:Defp}) we find (recall that for dust collapse $\Phi=0$, $e^{-\Psi} = \gamma$ and $t = \tau$)
$$
\pi^\tau = A,\quad
\gamma\pi^R = -A\cos\alpha,\quad
\ell = A r\sin\alpha,
$$
such that $\Pi = -\cos\alpha$. Introducing these relations into Eq.~(\ref{Eq:DefE}) and dividing by $\ell$ yields
\begin{equation}
\label{Eq:b_of_alpha}
b = b(\alpha,r)
 = \frac{r\sin\alpha}{\sqrt{1 + 2E_1} + \sqrt{\frac{2m_1}{r} + 2E_1}\cos\alpha},
\end{equation}
where we have used the equation $\gamma r' = \sqrt{1 + 2E_{1}}$, $E_1 := E(R_1)$, and the free-fall equation~(\ref{Eq:FreeFall}) evaluated at the surface of the cloud $R = R_1$. Note that for the particular values $\alpha = 0$ or $\alpha = \pi$ this formula gives $b = 0$, i.e. the correct impact parameter for ingoing and outgoing radial light rays. Inverting the relation~(\ref{Eq:b_of_alpha}) we obtain the two solutions
\begin{eqnarray}
\sin\alpha  &=& \frac{b r}{r^2 + b^2\left(\frac{2m_1}{r} + 2E_1\right)}\left[ 
\sqrt{1 + 2E_1} \pm \sqrt{\frac{2m_1}{r} + 2E_1}\sqrt{1 - \frac{b^2}{r^2}\left(1 - \frac{2m_1}{r}\right)} \right], 
\label{Eq:sin_alpha_of_b}\\
\cos\alpha &=& \frac{1}{r^2 + b^2\left(\frac{2m_1}{r} + 2E_1\right)}\left[ 
-b^2\sqrt{\frac{2m_1}{r} + 2E_1}\sqrt{1 + 2E_1} 
 \pm r^2\sqrt{1 - \frac{b^2}{r^2}\left(1 - \frac{2m_1}{r}\right)} \right],
\label{Eq:cos_alpha_of_b}
\end{eqnarray}
for $\alpha$ as a function of the impact parameter and the areal radius $r = r_1^+$ of the surface of the cloud at the event $x$. Evaluating Eqs.~(\ref{Eq:sin_alpha_of_b})~and~(\ref{Eq:cos_alpha_of_b}) at $b = 0$ yields $\sin\alpha = 0$, $\cos\alpha = \pm 1$ which shows that in this case it is the lower sign that determines the correct solution $\alpha = \pi$. More generally, it follows from Eq.~(\ref{Eq:drdlambda}) that
\begin{equation}
\frac{dr}{d\lambda}
 = A\frac{\sqrt{1 - \frac{b^2}{r^2}\left(1 - \frac{2m_1}{r} \right)}}{1 + \frac{b^2}{r^2}\left( \frac{2m_1}{r} + 2E_1 \right)}
 \left[ \mp\sqrt{1 + 2E_1} - \sqrt{\frac{2m_1}{r} + 2E_1}\sqrt{1 - \frac{b^2}{r^2}\left( 1 - \frac{2m_1}{r} \right)} \right].
\end{equation}
Since $\sqrt{2m_1/r + 2E_1} \leq \sqrt{1 + 2E_1}$ and $1 - (b/r)^2(1 - 2m_1/r) \leq 1$ it follows that the lower sign leads to an increase of $r$ along $\lambda$, and hence the corresponding light ray reaches the observer. Thus, the correct sign in Eqs.~(\ref{Eq:sin_alpha_of_b})~and~(\ref{Eq:cos_alpha_of_b}) is the lower one, and the incidence angle $\alpha$ is uniquely determined by these equations and lies in the interval\footnote{Notice that by restricting our considerations to the open interval $\pi/2 < \alpha < 3\pi/2$ rather to $\pi/2 \leq \alpha \leq 3\pi/2$ we have eliminated radiation that is incoming from infinity, grazes the surface of the collapsing cloud and returns back to infinity. For a discussion and an alternative description of this ``backward emitted radiation" see Ref.~\cite{jJ69}.} $\pi/2 < \alpha < 3\pi/2$. In particular, this implies that a segment of the null geodesic in the immediate past of $x$ lies inside the cloud, as required.

Now that we have managed to express the incidence angle $\alpha$ in terms of the impact parameter $b$ and the radius $r_1^+$ of the cloud when the null ray exits it, we need to establish the relation between $r_1^+$ and the proper time $\tau_{obs}$ at which the ray reaches the observer. We fix the time origin using the same normalization as in paper~I. We choose this  origin such that $\tau_{obs} = 0$ corresponds to the moment at which the observer encounters the radial light ray $\gamma_0$ that entered the cloud at the moment of time symmetry (see Fig.~\ref{Fig:Conformal}). In order to compute the relation between $\tau_{obs}$ and $r_1^+$ we work in standard Schwarzschild coordinates $(t,r)$ and first note that
$$
\tau_{obs} = \sqrt{1 - \frac{2m_1}{r_{obs}}}(t - t_0) \simeq t - t_0,
$$
where $t_0$ is the Schwarzschild time at $\tau_{obs} = 0$, and where in the last step we have used the assumption $r_{obs} \gg 2m_1$. Next, we use Eqs.~(\ref{Eq:DefE})~and~(\ref{Eq:EConsBis}), from which
$$
t - t_1^+ = \int\limits_{r_1^+}^{r_{obs}} \frac{dr}{N(r)\sqrt{1 - \frac{b^2}{r^2}N(r)}},
$$
where $t_1^+$ is the Schwarzschild time at which the light ray exits the cloud and where for notational simplicity we have set $N(r) := 1 - 2m_1/r$. Next, we use the fact that $r_1^+$ and $t_1^+$ are related to each other via the free-fall equations
$$
N(r)\dot{t} = {\cal E}_1 = \sqrt{1 + 2E_1},\qquad
\frac{1}{2}\dot{r}^2 - \frac{m_1}{r} = E_1,
$$
where we recall that the dot denotes differentiation with respect to the proper time of a free-falling observer comoving with the dust particles. From these two equations we find, upon integration from the moment of time symmetry, where $t = 0$ and $r_1 = R_1$, to the point where the cloud has collapsed to radius $r_1^+$, that
$$
t_1^+ = \sqrt{1 + 2E_1}\int\limits_{r_1^+}^{R_1} \frac{dr}{N(r)\sqrt{2E_1 + \frac{2m_1}{r}}}.
$$
Combining the above equations, we obtain the following expression
\begin{equation}
t(E_1,|b|,r_1^+) = \int\limits_{r_1^+}^{r_{obs}} \frac{dr}{N(r)\sqrt{1 - \frac{b^2}{r^2}N(r)}}
 + \sqrt{1 + 2E_1}\int\limits_{r_1^+}^{R_1} \frac{dr}{N(r)\sqrt{2E_1 + \frac{2m_1}{r}}}
\end{equation}
for the Schwarzschild time $t$ at the moment the null geodesic intersects the observer's world line, assuming that the geodesic has impact parameter $b$ and exited the cloud at radius $r_1^+$. By definition,
$$
t_0 = t(E_1,0,r_0),
$$
with $r_0$ the radius of the cloud at the moment the particular light ray $\gamma_0$ exits it. Hence, introducing the function
$$
F_b(r) := \frac{1}{\sqrt{1 - \frac{b^2}{r^2}N(r)}} - 1,
$$
we obtain
\begin{equation}
t(E_1,|b|,r_1^+) - t_0 = \int\limits_{r_1^+}^{r_{obs}} F_b(r) \frac{dr}{N(r)}
 - \int\limits_{r_0}^{r_1^+} \frac{dr}{N(r)}
 - \sqrt{1 + 2E_1}\int\limits_{r_0}^{r_1^+} \frac{dr}{N(r)\sqrt{2E_1 + \frac{2m_1}{r}}}.
\end{equation}
The last two integrals on the right-hand side can be computed analytically, see Appendix~C in~\cite{nOoS11}. The first integral converges as $r_{obs}\to \infty$, since $F_b(r)$ decays as $1/r^2$. Finally, we note that we can rewrite
$$
F_b(r) = \frac{b^2}{r^2}\frac{N(r)}
{\sqrt{1 - \frac{b^2}{r^2}N(r)}\left( 1 + \sqrt{1 - \frac{b^2}{r^2}N(r)} \right)},
$$
so that the first integral converges as $r_1^+\to 2m_1$.

Gathering the results and taking $r_{obs}\to \infty$ we finally obtain
\begin{equation}
\tau(E_1,|b|,r_1^+)
 = \int\limits_{r_1^+}^{\infty}\frac{b^2}{r^2}\frac{dr}
{\sqrt{1 - \frac{b^2}{r^2}N(r)}\left( 1 + \sqrt{1 - \frac{b^2}{r^2}N(r)} \right)}
 + 4m_1\log\left( \frac{U(y_0)}{U(y_1^+)} \right),
\label{Eq:tau}
\end{equation}
where
$$
U(y) = \frac{1}{a_1} \left( \sqrt{a_1^2-b_1^2y^2} - y\sqrt{1-b_1^2} \right) 
\exp \left\{ \frac{y^2}{2a_1^2} - \frac{\sqrt{1-b_1^2}}{2b_1^2}
\left[ \frac{1+2b_1^2}{b_1} \arctan \left( \frac{\sqrt{a_1^2-b_1^2y^2}}{b_1y} \right)
 +  \frac{y}{a_1^2}\sqrt{a_1^2-b_1^2y^2}    \right] \right\},
$$
with $a_1 := \sqrt{2m_1/R_1}$, $b_1 := \sqrt{-2E_1}$ and $y := \sqrt{r/R_1}$. In order to compute the remaining integral, it is convenient to introduce the new variable $s := |b|/r$ in terms of which
$$
\int\limits_{r_1^+}^{\infty}\frac{b^2}{r^2}\frac{dr}
{\sqrt{1 - \frac{b^2}{r^2}N(r)}\left( 1 + \sqrt{1 - \frac{b^2}{r^2}N(r)} \right)}
 = |b|\int\limits_0^{|b|/r_1^+}\frac{ds}
{\sqrt{1 - s^2  + \frac{2m_1}{|b|} s^3}\left( 1 + \sqrt{1 - s^2 + \frac{2m_1}{|b|} s^3} \right)}.
$$

The results of this subsection can be summarized as follows: if $\gamma$ is a null geodesic having an impact parameter $b$ restricted to $|b| < \bar{b}$, intersecting the observer's world line at time $\tau_{obs}\geq \tau_0$, then this geodesic $\gamma$ must intersect the cloud. The areal radius $r_1^+$ and the angle of incidence $\alpha$ [defined in Eq.~(\ref{Eq:Defalpha})] at the event $x$ at which the null geodesic exits the cloud can be determined from $\tau_{obs}$ and $b$ by first inverting the equation $\tau(E_1,|b|,r_1^+) = \tau_{obs}$ to find $r_1^+$ and then using Eqs.~(\ref{Eq:sin_alpha_of_b})~and~(\ref{Eq:cos_alpha_of_b}) with $r = r_1^+$ to determine $\alpha$. Given $r_1^+$ and $\alpha$ we can continue tracing back the null geodesic into the cloud, and this procedure is described next.

\subsection{Step 2: Tracing back the null rays inside the cloud and determining the critical angle $\hat{\alpha}$}
\label{Sec:Step2}

Starting from the surface of the cloud at the event $x$ with $r = r_1^+$, at first we introduce the special angle $\bar{\alpha} := \alpha(\bar{b},r_1^+)$ obtained from Eqs.~(\ref{Eq:sin_alpha_of_b})~and~(\ref{Eq:cos_alpha_of_b}) and subject to the constraint  $\bar{\alpha} \in (\pi/2,\pi]$. As explained in the previous subsection, the light rays with incidence angle $\alpha\in (\bar{\alpha}, 2\pi - \bar{\alpha})$ are those that have large enough energy to make it over the potential barrier, see Eq.~(\ref{Eq:EConsBis}). Subsequently, for any $\alpha\in (\bar{\alpha}, 2\pi - \bar{\alpha})$, we solve numerically the dynamical system in Eq.~(\ref{Eq:DynSys}) in order to trace the nonradial null geodesic backwards into the dust cloud and determine whether it exits the cloud in the past or hits the singularity. If the collapse results into a black hole, only the former possibility can occur, whereas in the case of a naked singularity formation, the latter possibility may occur as soon as the event $x$ lies to the future of the Cauchy horizon and the impact parameter $b$ is sufficiently small in magnitude. As we have already discussed in the Introduction and have shown in detail in Sec.~\ref{Sec:rays_from_sing}, there exist future directed null geodesics emanating from the naked singularity, so it is possible that one of these geodesics intersects the surface of the cloud at the event $x$ with incidence angle $\alpha$ within the admissible range $(\bar{\alpha}, 2\pi - \bar{\alpha})$ of the observer's field of view. To deal with this new possibility, we introduce a critical angle $\hat{\alpha}$ defined in such a way that null geodesics traced back from $x$ hit the singularity in their past if $\alpha \in (\hat{\alpha},2\pi - \hat{\alpha})$ and exit the cloud in their past otherwise. If $x$ lies to the past of the Cauchy horizon (in particular, this is always true in the black hole case), then $\hat{\alpha} = \pi$ and the light ray always exits the cloud in its past. However, as we will show below, as soon as the event $x$ at the cloud's surface crosses the Cauchy horizon, $\hat{\alpha}$ decreases from $\pi$ to a value smaller than $\bar{\alpha}$ as the radius of the cloud decreases to $2m_1$, implying that there are more and more light rays in the observer's field of view that emanate from the singularity instead of the distant source. Geometrically, the angle $\hat{\alpha}$ determines the optical size of the naked singularity as seen by a free-falling observer which is comoving with the surface of the cloud. An analytic expression for $\hat{\alpha}$ in the particular case of a self-similar Tolman-Bondi cloud is derived in Appendix~\ref{App:SS}.

\begin{figure}[h!]
\begin{center}
\includegraphics[width=6.35cm]{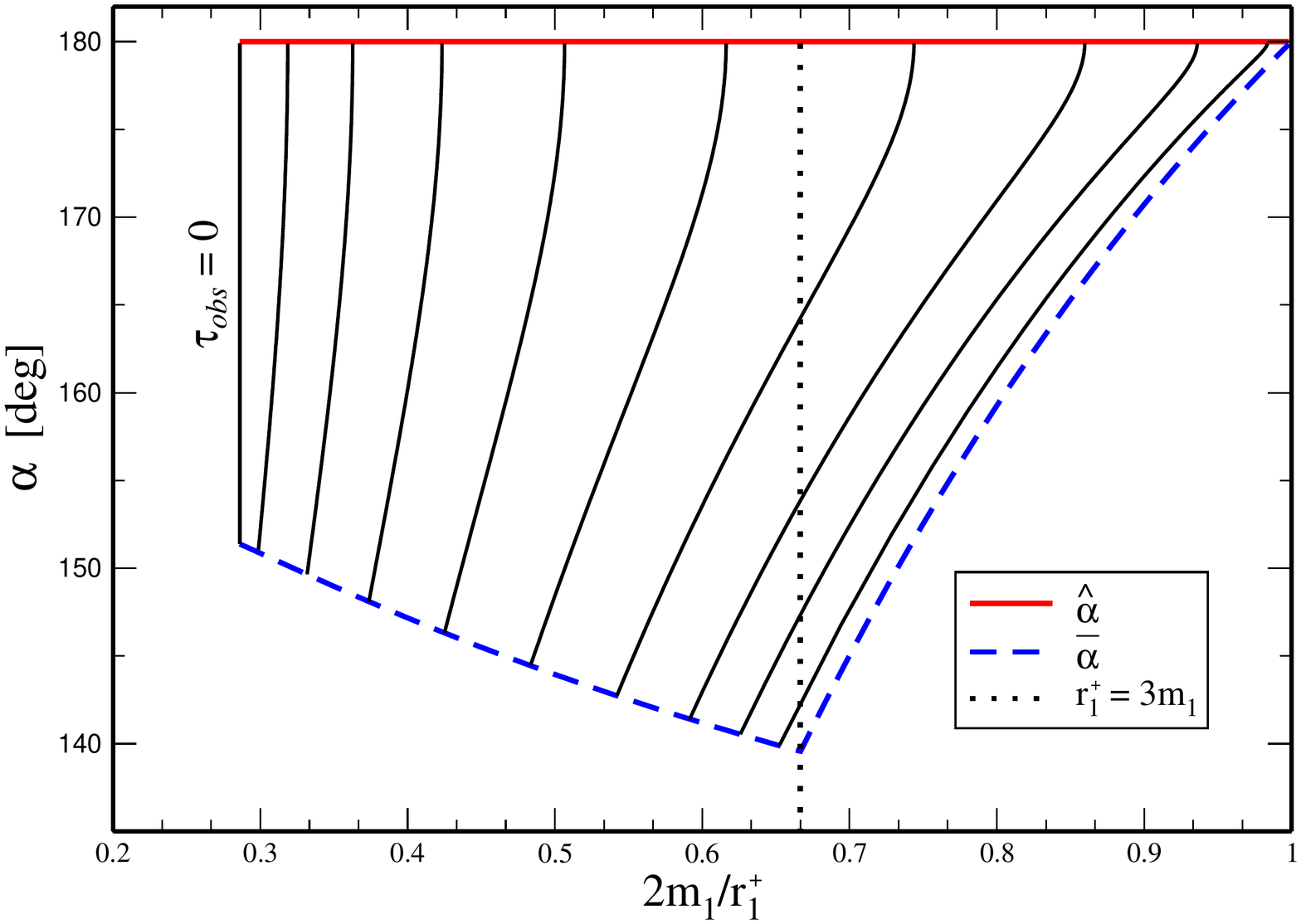}
\hspace{-0.8cm}
\includegraphics[width=6.35cm]{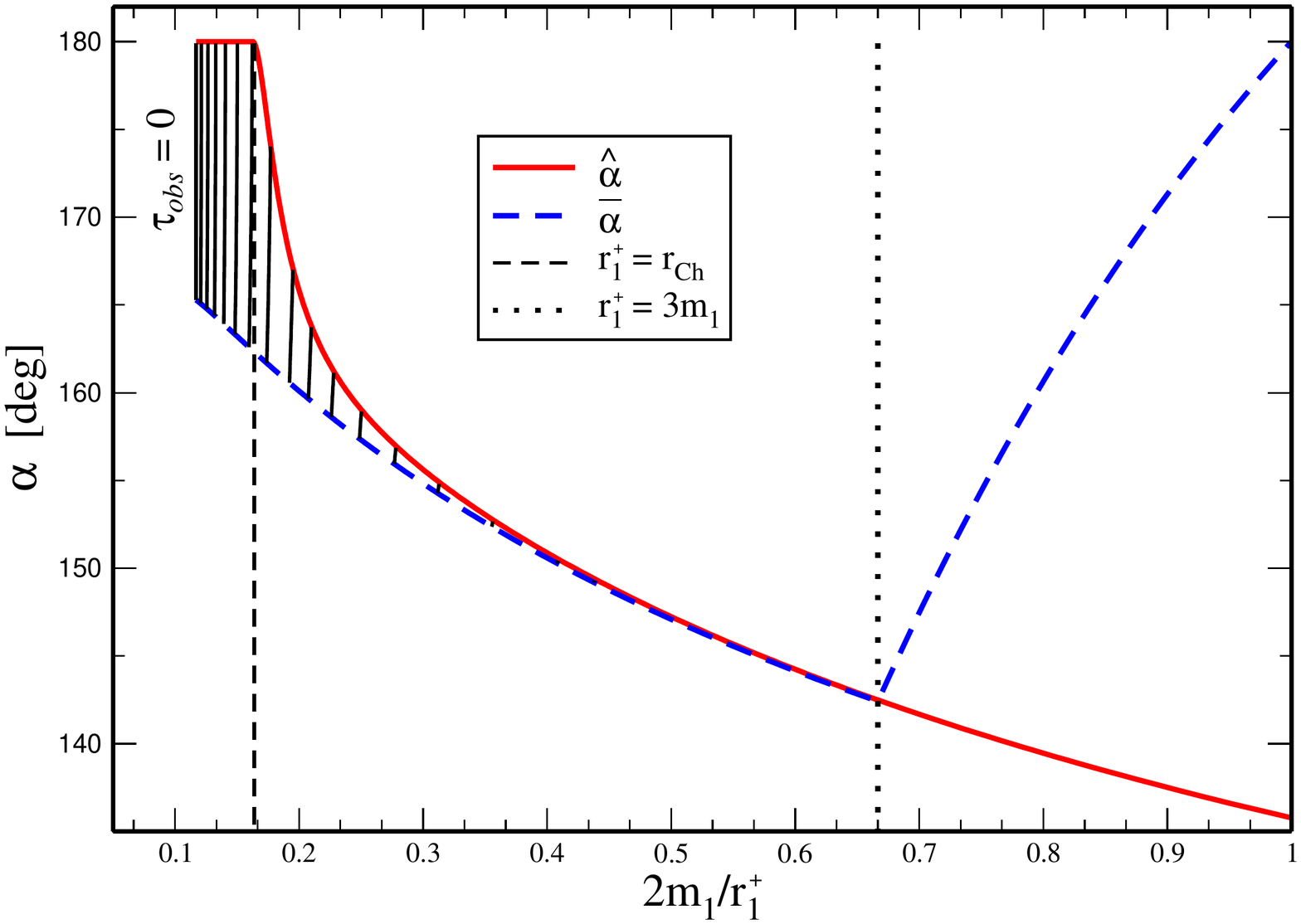}
\hspace{-0.8cm}
\includegraphics[width=6.35cm]{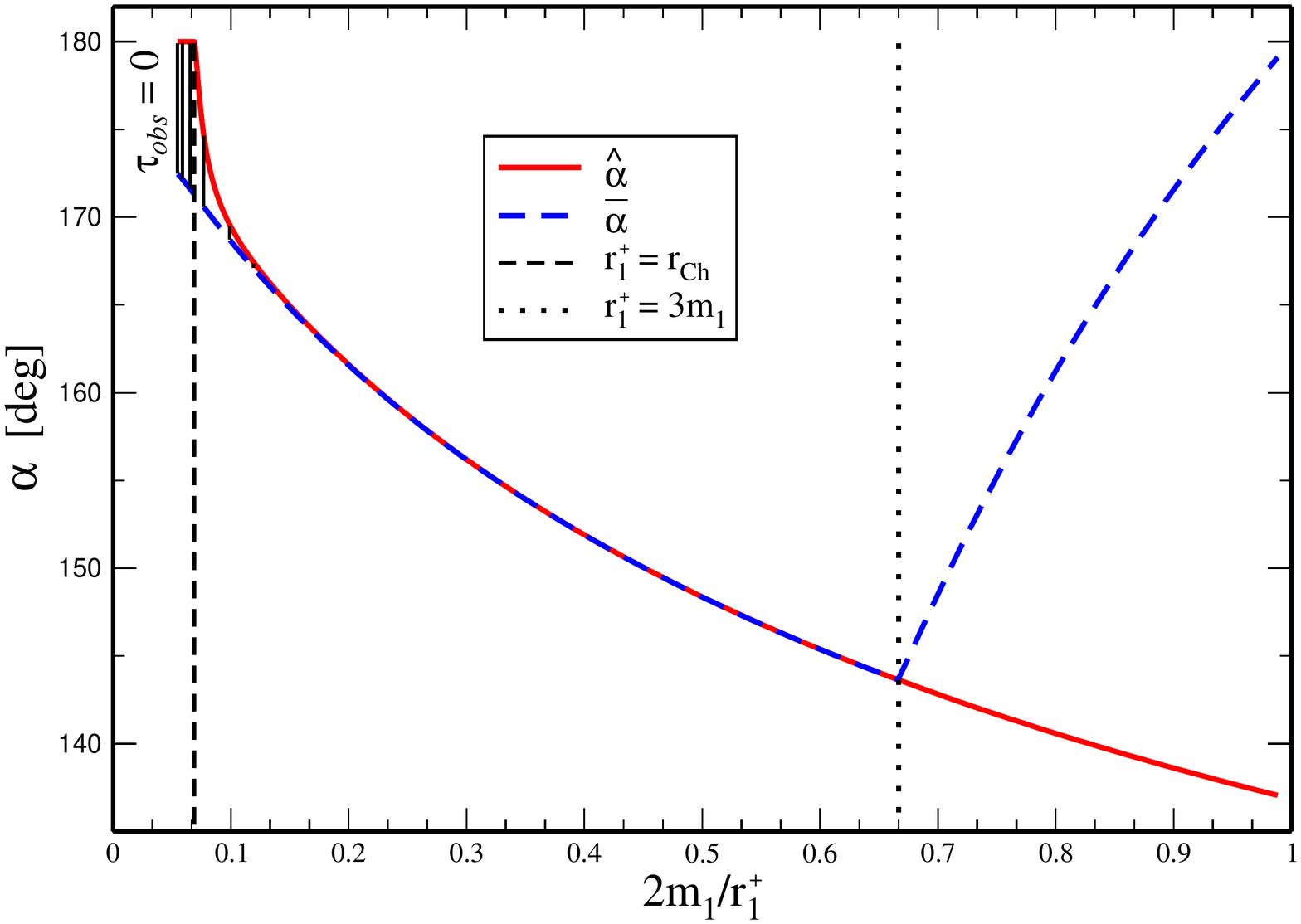}
\end{center}
\caption{\label{Fig:Alphas}
The critical angle $\hat{\alpha}$ (thick solid lines) as well as $\bar{\alpha}$ (thick dashed lines) as functions of $2m_1/r_1^+$. The thin solid lines correspond to slices of constant proper time of the observer. Left panel: Black hole case with initial compactness ratio ${\cal C} = {\cal C}_0$. Central panel: Naked singularity with ${\cal C} = {\cal C}_0/2$. Right panel: Naked singularity with ${\cal C} = {\cal C}_0/4$. $r_{Ch}$ denotes the areal radius of the surface of the star at the moment it crosses the Cauchy horizon. Note that in the black hole case $\hat{\alpha} = \pi$ since the asymptotic observer always lies to the past of the Cauchy horizon in this case.}
\end{figure}

In the generic, bounded Tolman-Bondi collapse, an analytic determination of the critical angle $\hat{\alpha}$ is probably not feasible. Therefore, we determine $\hat{\alpha}$ numerically by bisection in $\alpha$ for every event $x$ at the surface of the star such that $\tau_{obs} \geq 0$ and $r_1^+ > 2m_1$. Figure~\ref{Fig:Alphas} shows $\hat{\alpha}$ and $\bar{\alpha}$ as functions of $2m_1/r_1^+$ for three different initial compactness ratios ${\cal C}$ (see the discussion at the end of Sec.~\ref{Sec:ID} for the precise definition of ${\cal C}$). In the left panel, ${\cal C} = {\cal C}_0$, and thus a black hole appears, whereas the central and right panels correspond to the cases of naked singularities arising from collapsing clouds with ${\cal C} = {\cal C}_0/2$ and ${\cal C} = {\cal C}_0/4$, respectively. For every given $\tau = \tau_{obs}$ and a discrete set of equally spaced values of the impact parameter $b\in [0,\bar{b})$, we numerically invert Eq.~(\ref{Eq:tau}) to find $r_1^+$ and then, from Eqs.~(\ref{Eq:sin_alpha_of_b})~and~(\ref{Eq:cos_alpha_of_b}) with $r = r_1^+$, we obtain discrete sets of $\alpha \in (\bar{\alpha},\hat{\alpha})$ producing the thin solid lines in each panel of Fig.~\ref{Fig:Alphas}. Notice that the dependency of $\bar{\alpha}$ on $2m_1/r_1^+$ shows a kink at $2m_1/r_1^+ = 2/3$, corresponding to the moment the cloud's surface coincides with the photosphere. The origin of this kink is related to the fact that the function $1 - \bar{b}^2(1 - 2m_1/r)/r^2\geq 0$, appearing on the right-hand side of Eqs.~(\ref{Eq:sin_alpha_of_b})~and~(\ref{Eq:cos_alpha_of_b}), has a second-order root at $r = 3m_1$, implying a change of sign in the derivative of its square root.

\begin{figure}[h!]
\begin{center}
\includegraphics[width=7.cm]{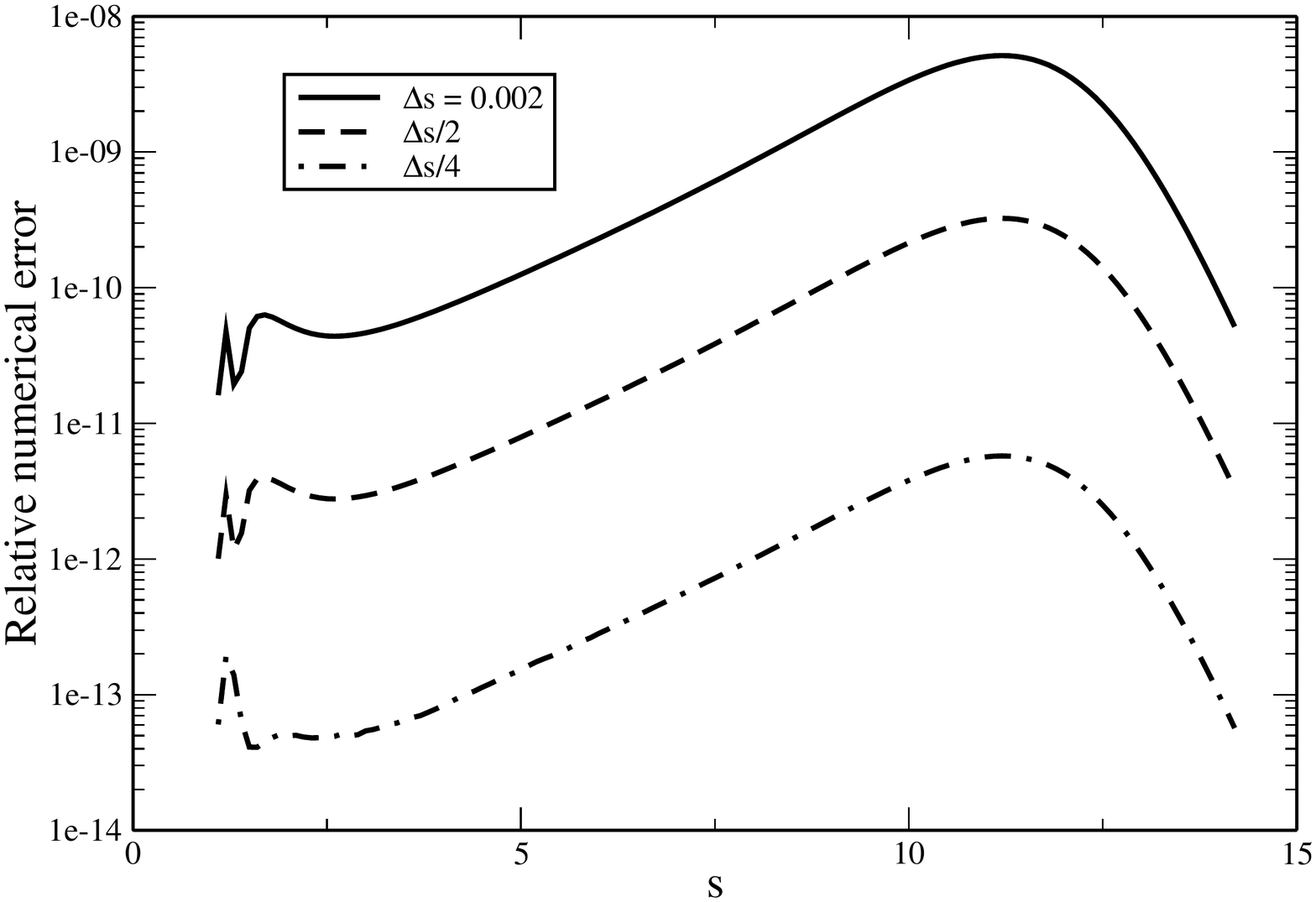}
\hspace{-0.1cm}
\includegraphics[width=7.cm]{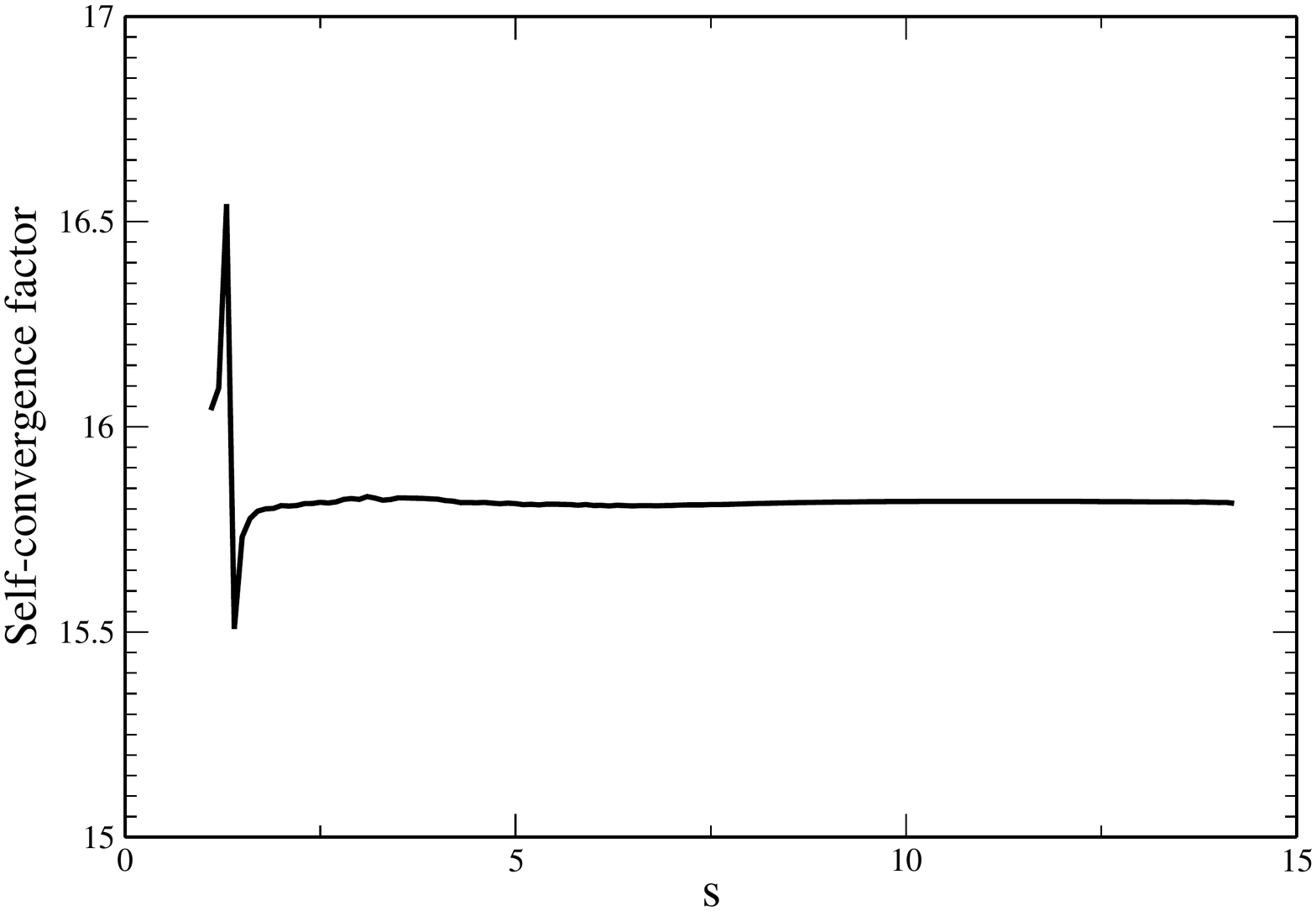}
\end{center}
\caption{\label{Fig:Convergence} Self-convergence test for the numerical solution of the dynamical system in Eq.~(\ref{Eq:DynSys}) for a null geodesic traced back from the event $x$ at the surface of the cloud, passing very close to the singularity, and intersecting again the surface of the cloud in the past. The relative numerical error decreases (left panel) by a factor of $\sim16$ (right panel) when the step size decreases by a factor of $2$.}
\end{figure}

In order to demonstrate the accuracy of our numerical method, in Fig.~\ref{Fig:Convergence} we present a self-convergence test for the numerical solution of the dynamical system in Eq.~(\ref{Eq:DynSys}) computed by a fourth-order Runge-Kutta method with step size $\Delta s = 0.002$, for a null geodesic traced back from the event $x$ until it intersects again the cloud's surface in the past. In Fig.~\ref{Fig:Convergence} we show results for an incidence angle $\alpha$ lying very close to the critical value $\hat{\alpha}$ such that the corresponding null geodesic grazes the singularity. The plot shows that the relative numerical error decreases (left panel) by a factor $\sim16$ (right panel) when the step size decreases by a factor $2$.
%

\subsection{Step 3: Measuring the redshift of the photons originating from the source}

Once we have a light ray that exits the collapsing star in the past, we know that it must extend all the way to the past asymptotic region, where the corresponding photon is emitted with frequency $\nu_\infty^-$. In this step, we evaluate the total gravitational frequency shift $\nu_\infty^+/\nu_\infty^-$ measured by the distant observer in the future asymptotic region, where $\nu_\infty^+$ denotes the detected frequency. This frequency shift is given by the simple formula~(\ref{Eq:Total_redshift}) which only requires the computation of the quantity ${\cal E}$ along the portion of the light ray inside the star.

We numerically evaluate the frequency shift perceived by the observer at different proper times $\tau_{obs} > 0$ and for different impact parameters $b\in (0,\bar{b})$ of the light ray.\footnote{Because of spherical symmetry it is sufficient to consider positive values of $b$, corresponding to the interval $\alpha\in (\pi/2,\pi]$ for the incidence angle.} To  this purpose, we first determine the angle of incidence $\alpha\in (\bar{\alpha},\pi)$ for each chosen value of $\tau_{obs}$ and impact parameter $b$. In the naked singularity case we check whether $\alpha < \hat{\alpha}$ or $\alpha > \hat{\alpha}$. In the black hole case or in the naked singularity case with $\alpha < \hat{\alpha}$, the light ray extends all the way to past null infinity according to the definition of $\hat{\alpha}$, and the total redshift is computed using Eq.~(\ref{Eq:Total_redshift}). In the naked singularity case with $\alpha > \hat{\alpha}$, the light ray originates from the naked singularity and there is an infinite redshift, making it practically invisible.

\begin{figure}[h!]
\begin{center}
\includegraphics[width=6.3cm]{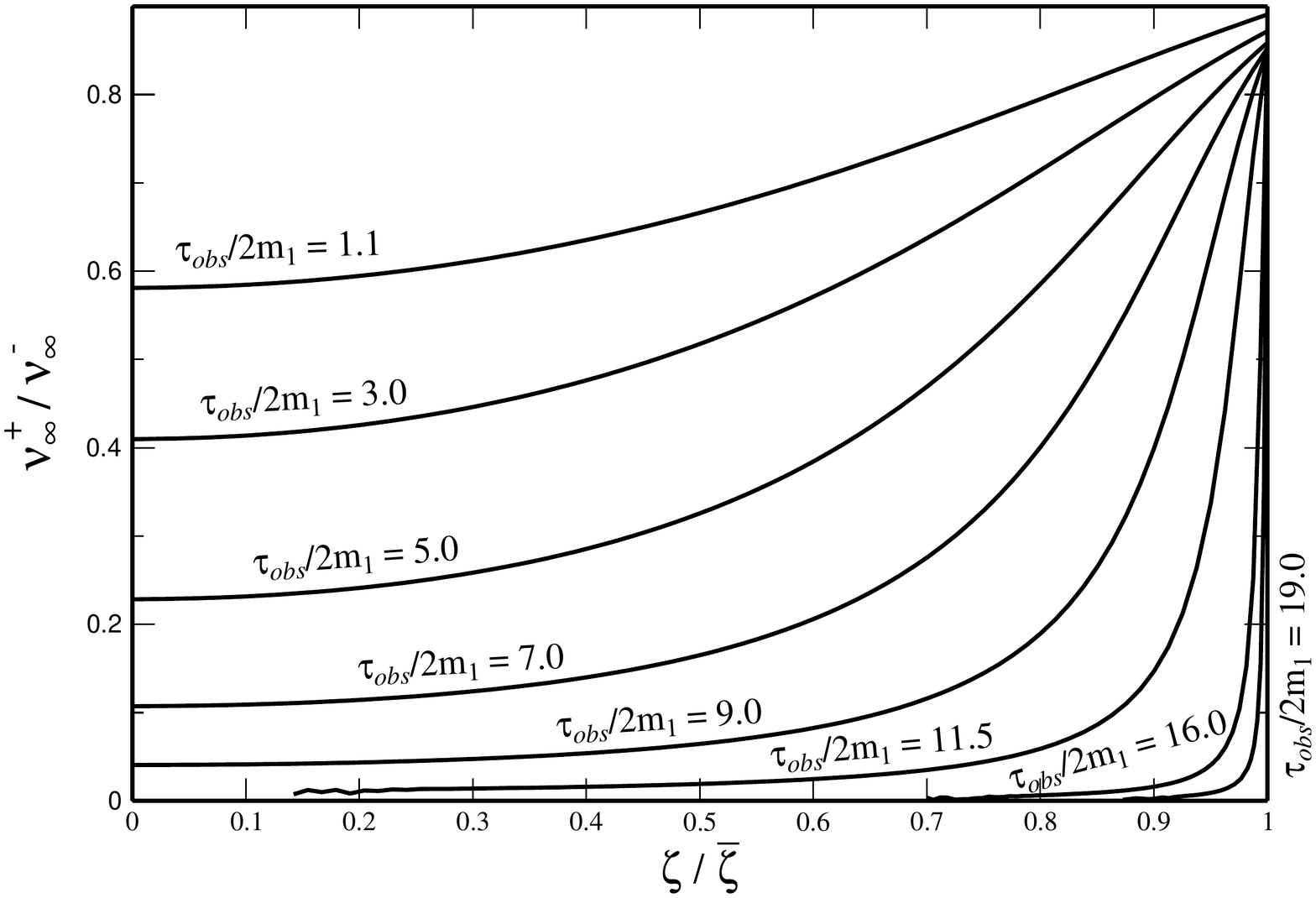}
\hspace{-0.7cm}
\includegraphics[width=6.3cm]{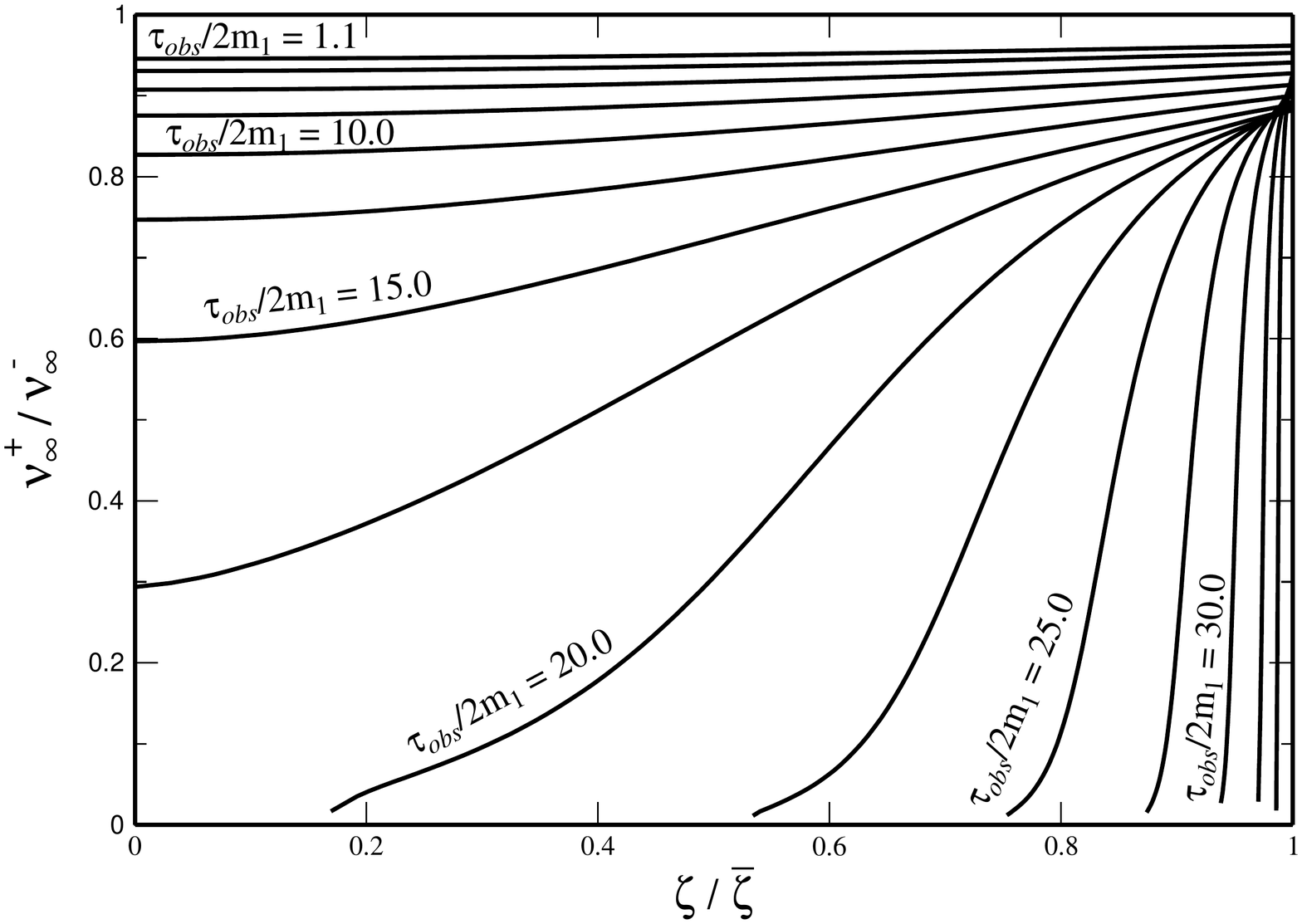}
\hspace{-0.7cm}
\includegraphics[width=6.3cm]{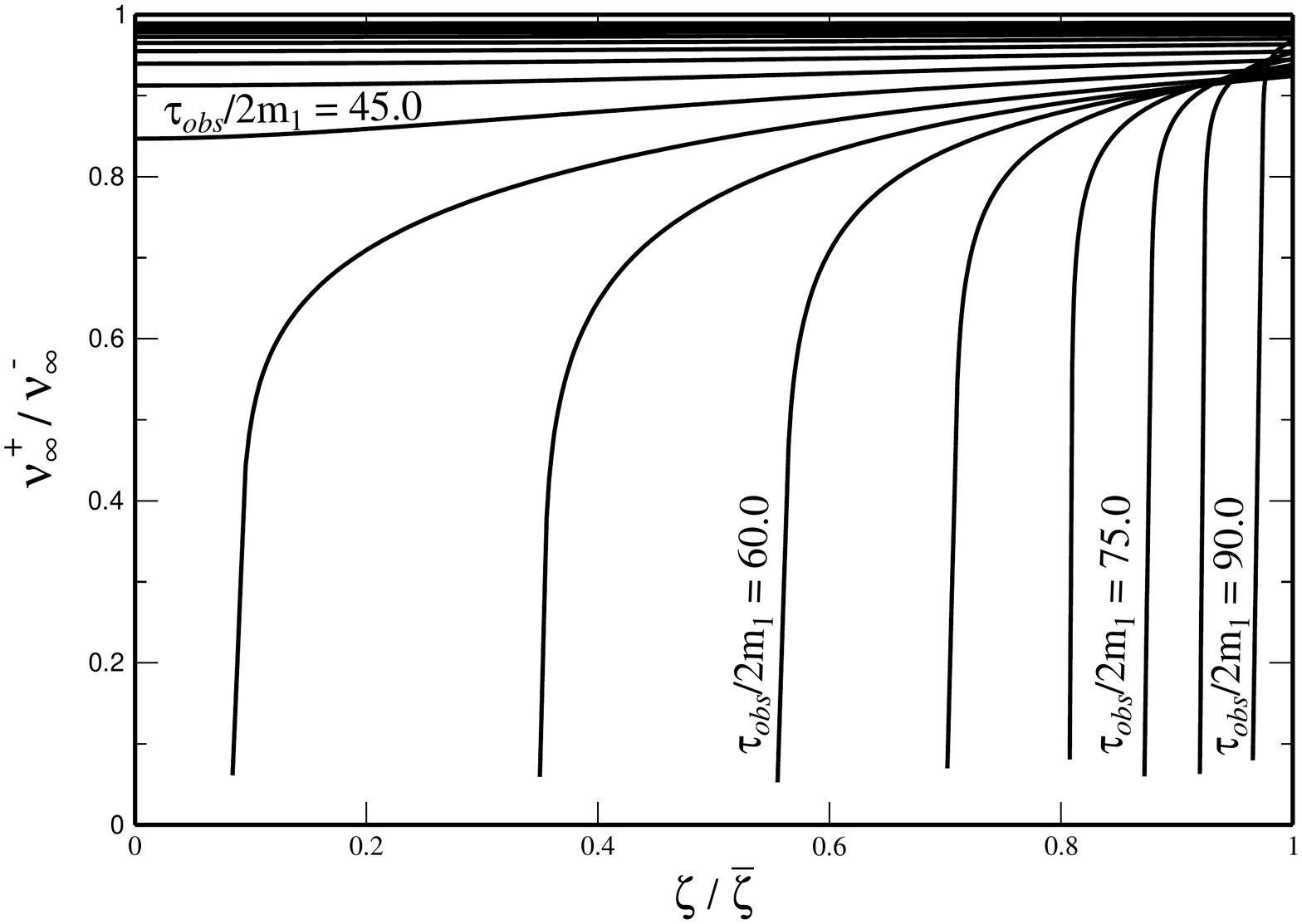}
\end{center}
\caption{\label{Fig:Redshift}
Total redshift measured by the asymptotic observer at certain proper times as a function of the normalized viewing angle $\zeta/\bar{\zeta}$ of the collapsing object. Left panel: Black hole case with initial compactness ratio ${\cal C} = {\cal C}_0$, for which $\tau_{3m_1}/2m_1 \approx 10.8$. Central panel: Naked singularity with ${\cal C} = {\cal C}_0/2$. In this case $\tau_{Ch}/2m_1 \approx 18.4$ and $\tau_{3m_1}/2m_1 \approx 41.1$. Right panel: Naked singularity with ${\cal C} = {\cal C}_0/4$, for which $\tau_{Ch}/2m_1 \approx 47.5$ and $\tau_{3m_1}/2m_1 \approx 121.9$.}
\end{figure}

For the same cases as the ones presented in Fig.~\ref{Fig:Alphas}, in Fig.~\ref{Fig:Redshift} we show the total redshift $\nu_\infty^+/\nu_\infty^-$ as a function of the viewing angle $\zeta$ (see definition below) as seen by the asymptotic observer for certain fixed values of $\tau_{obs} > 0$. We denote by $\tau_{Ch}$ ($\tau_{3m_1}$) the proper time at which the observer starts detecting light rays that exit the cloud inside the Cauchy horizon (photosphere). The viewing angle $\zeta$ is defined\footnote{Here, a viewing direction is a spatial direction in the observer's rest frame that describes the spatial direction of a photon originating from the source that has traversed the cloud and has been detected by the observer. It is related but not identical to the celestial coordinates of the observer's sky.} in a similar way as the angle of incidence $\alpha$, see Eq.~(\ref{Eq:Defalpha}). To define $\zeta$, we choose the orthonormal tetrad
$$
{\bf e}_0 = \frac{1}{\sqrt{N(r_{obs})}}\frac{\partial}{\partial t},\quad
{\bf e}_1 = \sqrt{N(r_{obs})}\frac{\partial}{\partial r},\quad
{\bf e}_2 = \frac{1}{r}\frac{\partial}{\partial\vartheta},\quad
{\bf e}_3 = \frac{1}{r\sin\vartheta}\frac{\partial}{\partial\varphi}
$$
adapted to the static observer, and we parametrize the photon's four-momentum as
\begin{equation}
{\bf p} = B\left( {\bf e}_0 + \cos\zeta\,{\bf e}_1 + \sin\zeta\, {\bf e}_3 \right),
\label{Eq:Defzeta}
\end{equation}
with $B > 0$ a positive constant. Using Eqs.~(\ref{Eq:DefE})~and~(\ref{Eq:ECons}) we find
$$
\sin\zeta = \frac{b}{r_{obs}}\sqrt{1 - \frac{2m_1}{r_{obs}}},
$$
the angle $\zeta = 0$ corresponding to photons with zero impact parameter (radial light rays). Since $r_{obs} \gg 2m_1,b$, this relation simplifies to
\begin{equation}
\zeta = \frac{b}{r_{obs}}.
\end{equation}
In particular, the largest viewing angle allowed by our assumption ($|b| \leq \bar{b}$) is $\bar{\zeta} := \bar{b}/r_{obs}$.

A comparison between the different panels in Fig.~\ref{Fig:Redshift} reveals that  the redshift perceived by the asymptotic observer depends crucially upon the collapse outcome. In particular, regarding the total redshift suffered by radial photons (corresponding to $\zeta=0$) as a function of proper time, it can be seen from Fig.~\ref{Fig:Redshift} that our previous results in paper~I are recovered. Namely, for the case of an event horizon formation, the left panel in Fig.~\ref{Fig:Redshift} shows that in an amount of proper time of the order $10m_{1}$ crossing times, the observer finds that the ratio $\nu_\infty^+/\nu_\infty^-$ smoothly approaches zero. In contrast to this, for the naked singularity case, the central and right panels in Fig.~\ref{Fig:Redshift} show a sharp cutoff of the signal from the radial direction, which can occur even at crossing times of the order $100m_1$, depending on the initial compactness ratio of the collapsing object. For a clearer picture of this behavior see also Figs. 2 and 3 in paper I.

Furthermore, Fig.~\ref{Fig:Redshift} also shows that the behavior of the total redshift of radial photons persists for nonradial ones. That is, for every fixed viewing angle $\zeta \ne 0$, the redshift function decays smoothly to zero in the case of an event horizon formation, whereas in the naked singularity case the decay is slower initially but then shows a sharp cutoff to zero after the observer crosses the Cauchy horizon. Nevertheless, we see that as $\zeta$ approaches $\bar{\zeta}$, the redshift seems to be independent of the collapse time. This can be understood by noting that  $\zeta=\bar{\zeta}$ implies that we are dealing with null geodesics originating at the photosphere, and thus the associated redshift is independent of the collapse (for a related discussion of this point see Ref.~\cite{wAkT68} and in particular Fig.~2 of that paper).

In addition, Fig.~\ref{Fig:Redshift} shows that in the naked singularity case the redshift behavior of nonradial photons as a function of $\zeta$ for late $\tau_{obs}$ is also remarkable in the following sense: once the observer crosses the Cauchy horizon ($\tau_{obs} > \tau_{Ch}$), every light ray suddenly experiences an infinite redshift at some viewing angle $\zeta > 0$, giving rise to a dark disk which will be discussed in detail in the next section.

\section{The shadow of the naked singularity compared to the shadow of the black hole}
\label{Sec:Results}

In this section, we provide a description of the ``optical" appearance of the collapsing cloud as perceived in the eyes of an asymptotic observer, based on the implications of the graphs in Figs.~\ref{Fig:Alphas}~and~\ref{Fig:Redshift}. We first examine the behavior of the angles $\bar{\alpha}$ and $\hat{\alpha}$ as a function of $2m_1/r_1^+$. First, the panels in Fig.~\ref{Fig:Alphas} show that the angle $\bar{\alpha}$ exhibits a qualitative behavior which is independent of whether an event horizon or a Cauchy horizon forms. This property can be understood by noting that as long as the surface of the cloud is in free fall, its behavior is entirely determined by the exterior Schwarzschild geometry.

Next, we emphasize that in the case of an event horizon formation (see left panel of Fig.~\ref{Fig:Alphas}), any $\alpha \in (\bar{\alpha},\pi]$ [see Eq.~(\ref{Eq:Defalpha}) for the definition of $\alpha$] defines a future directed null geodesic exiting the cloud and reaching the asymptotic observer, while in the past direction it traverses the cloud and reaches the source. Even though the observer is always allowed to receive photons from every direction such that $\alpha \in (\bar{\alpha},\pi]$, after the surface of the cloud crosses inward the photosphere at $r = 3m_1$, the bright directions in the observer's sky progressively reduce due to the high redshift experienced by radial and nearly radial photons (see left panel of Fig.~\ref{Fig:Redshift}). In contrast to this, when a Cauchy horizon forms (see the central and right panels of Figs.~\ref{Fig:Alphas}~and~\ref{Fig:Redshift}), the bright directions in the observer's sky are determined by those values of $\alpha$ lying in the interval $(\bar{\alpha},\hat{\alpha})$. Clearly, there is a dramatic reduction in the bright directions in the observer's sky and the interval $(\bar{\alpha},\hat{\alpha})$ shrinks to a point as the cloud's surface reaches the photosphere.\footnote{It may be noticed that the central and right panels of Fig.~\ref{Fig:Alphas} show that after the cloud's surface crosses the photosphere $\hat{\alpha} < \bar{\alpha}$. This reversal in magnitude is irrelevant from the point of view of the asymptotic observer since the trajectories of null geodesics with impact parameter $b$ larger than $\bar{b}$ lie entirely in the Schwarzschild region a soon as $r_1^+ < 3m_1$.}

\begin{figure}[h!]
\begin{center}
\includegraphics[width=18.4cm]{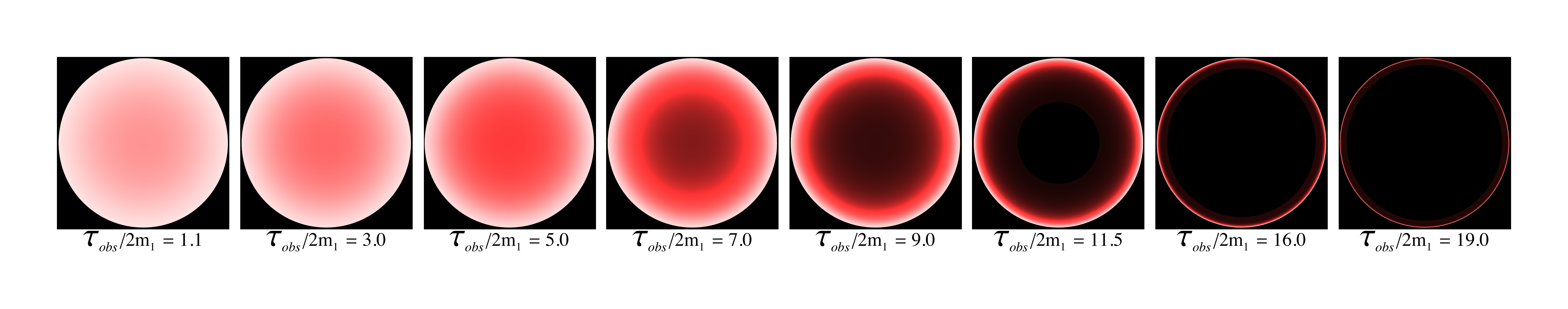}
\end{center}
\caption{\label{Fig:BH_snapshots} Redshift snapshots at different proper times of the asymptotic observer as the dust star of initial compactness ratio ${\cal C}_0$ collapses to a black hole. Here, $\tau_{3m_1}/2m_1 \approx  10.8$. 
}
\end{figure}

\begin{figure}[h!]
\begin{center}
\includegraphics[width=18.4cm]{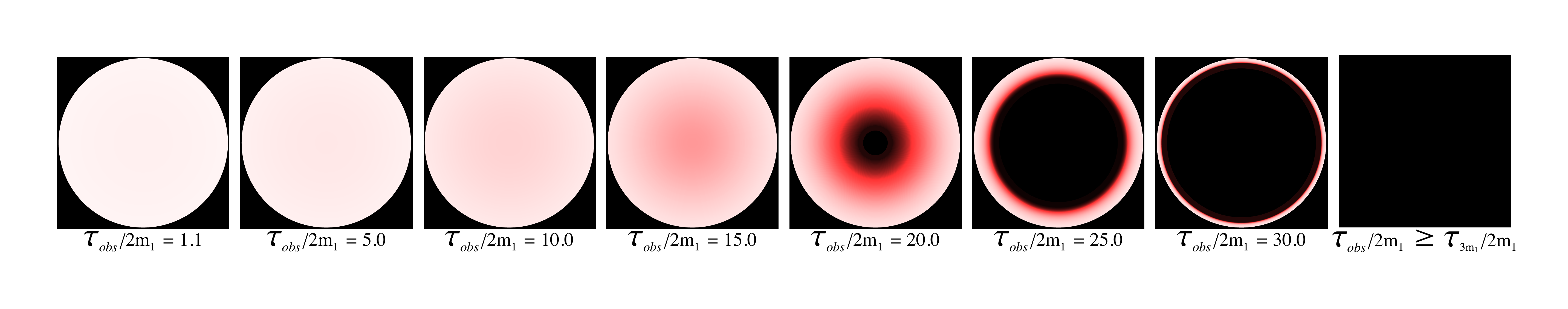}
\end{center}
\caption{\label{Fig:N05_snapshots} Redshift snapshots at different proper times of the asymptotic observer as the dust star of initial compactness ratio ${\cal C}_0/2$ collapses to a naked singularity. In this case, $\tau_{Ch}/2m_1 \approx 18.4$, and $\tau_{3m_1}/2m_1 \approx 41.1$. The shadow of the naked singularity appears at $\tau_{obs} = \tau_{Ch}$, and then it covers the whole region at $\tau_{obs} = \tau_{3m_1}$.}
\end{figure}

\begin{figure}[h!]
\begin{center}
\includegraphics[width=18.4cm]{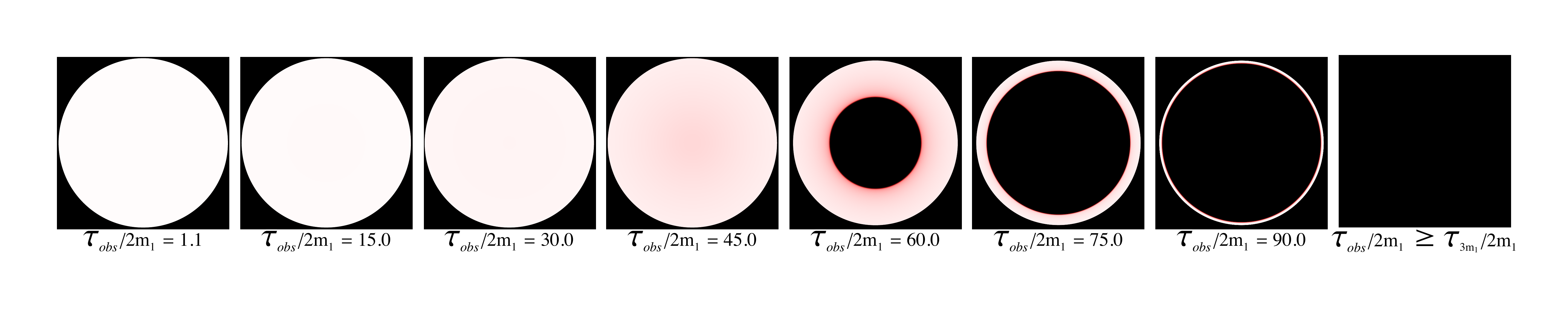}
\end{center}
\caption{\label{Fig:N025_snapshots} Redshift snapshots at different proper times of the asymptotic observer as the dust star of initial compactness ratio ${\cal C}_0/4$ collapses to a naked singularity. Here, $\tau_{Ch}/2m_1 \approx 47.5$, and $\tau_{3m_1}/2m_1 \approx 121.9$. The same qualitative features as in Fig.~\ref{Fig:N05_snapshots} are present.}
\end{figure}

In order to provide the reader with an image of the observer's view as he or she moves to the future, in Figs.~\ref{Fig:BH_snapshots},~\ref{Fig:N05_snapshots}~and~\ref{Fig:N025_snapshots} we show color maps of evenly selected curves from Fig.~\ref{Fig:Redshift}, where the colors indicate the redshift's magnitude, the darker regions corresponding to higher redshifts. Figure~\ref{Fig:BH_snapshots} corresponds to the case of black hole formation. In this case, we see that the inner zone of the object progressively becomes darker, indicating a high redshift for nearly radial light rays. When the observer starts detecting light rays that exit the cloud inside the photosphere (that is, for $\tau_{obs} > \tau_{3m_1}$ and $r_1^+ < 3m_1$), there is already a high redshift zone at the center, and this zone becomes fainter and grows continuously until it covers nearly the whole region under consideration $\zeta < \bar{\zeta}$ as $\tau_{obs}$ becomes large and $r_1^+ \to 2m_1$. These features show the dynamical formation of the {\it shadow of the black hole}, the final image consisting of a black region surrounded by a bright light ring due to the strong lensing effect at the photosphere.

When a naked singularity results from the dust collapse instead of a black hole, we observe a completely different behavior of the shadow. This is mainly due to the nontrivial dependency of the critical angle $\hat{\alpha}$ on the compactness ratio $2m_1/r_1^+$; see Fig.~\ref{Fig:Alphas}. As expected, $\hat\alpha$ becomes strictly smaller than $\pi$ from the moment the observer starts detecting light rays that exit the cloud inside the Cauchy horizon (which corresponds to $\tau_{obs} > \tau_{Ch}$ and $r_1^+ < r_{Ch}$). Moreover, the region $\alpha > \hat\alpha$ increases monotonously as the cloud collapses, and as a consequence, for $\tau_{obs} > \tau_{Ch}$, the observer sees a dark disk concentric with the image of the cloud which occults the light rays coming from the source, see Figs.~\ref{Fig:N05_snapshots}~and~\ref{Fig:N025_snapshots}. This disk grows continuously until it covers the whole region under consideration at finite time $\tau_{obs} = \tau_{3m_1}$ (corresponding to $r_1^+ = 3m_1$) which remarkably coincides with the intersection of $\hat{\alpha}$ and $\bar{\alpha}$; see the central and right panels of Fig.~\ref{Fig:Alphas}. This effect is also evident in the central and right panels of Fig.~\ref{Fig:Redshift}, since there is a cutoff in the redshift function at some $\zeta > 0$ for each $\tau_{obs} > \tau_{Ch}$, similar to the effect reported before in the case of radial light rays in paper~I. Indeed, the increasing opaque disk arising in the naked cases corresponds to null rays coming precisely from the naked singularity, which are infinitely redshifted as discussed in Sec.~\ref{Sec:Infinite_redshift}. In this sense, we call this disk {\it the shadow of the naked singularity}. 

\begin{figure}[h!]
\begin{center}
\includegraphics[width=9.5cm]{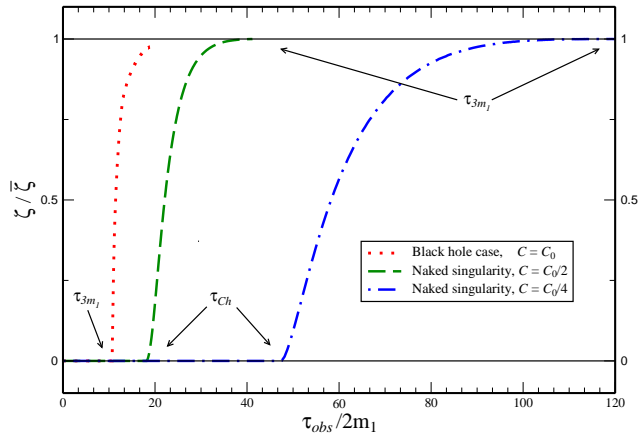}
\end{center}
\caption{\label{Fig:Shadows} Normalized angular extension of the shadow of the black hole (dotted line) compared to the shadow of two naked singularities (dashed lines), as functions of the observer's proper time. In the black hole case, the shadow becomes apparent around $\tau_{obs} = \tau_{3m_1}$ ($r_1^+ = 3m_1$), and then it grows continuously and takes an infinite time to cover the object, leaving a thin bright ring around it. In the naked singularity cases, the shadow appears in $\tau_{obs} = \tau_{Ch}$, and then it grows continuously and covers the whole region under consideration in finite time $\tau_{obs} = \tau_{3m_1}$. 
}
\end{figure}

In order to compare the dynamical formation of the different shadows discussed so far, in Fig.~\ref{Fig:Shadows} we plot their angular extension as a function of the proper time of the asymptotic observer. The dotted line corresponds to the shadow of the black hole, while the dashed lines correspond to the shadows of the naked singularities. An evident difference is the time scale in which the shadows grow. It turns out that 97\% of the shadow of the black hole develops in an interval of $\sim16.2m_1$ crossing times, whereas the same growth of the shadow of the naked singularity with initial compactness ratio ${\cal C} = {\cal C}_0/2$ takes $\sim45.4m_1$ crossing times, and $\sim148.8m_1$ crossing times for the naked singularity with the smallest initial compactness ratio ${\cal C} = {\cal C}_0/4$. This means that the shadow of these naked singularities takes around 2.8 and 9.1 times longer to reach the same angular extension of the shadow of the black hole.

Regarding the final state of the collapsing object and its appearance with respect to asymptotic observers, we note that a bright ring remains at late times in the black hole case, whereas the shadow of the naked singularity covers the whole region in finite time $\tau_{3m_1}$. Although the nature of the shadow in the naked singularity case is essentially different from the black hole case, since it is due to infinitely redshifted photons emanating from the singularity instead of highly redshifted photons originating from the illuminating source, it is important to stress that there is no difference in the image of the source for light rays with impact parameters larger than $\bar{b}$ at late times. This is due to the fact that at late times, light rays with $b > \bar{b}$ are scattered at the potential barrier, and thus they propagate entirely in the Schwarzschild exterior region. This suggests that observing the final state of the source's image is insufficient to distinguish a naked singularity from a black hole. In this sense, the singularity remains censored by its own shadow and moreover, it could mimic a black hole shadow. However, as we have discussed, the dynamical behavior of the redshift during the formation of the shadow represents a possibility to distinguish between the two cases.

\section{Conclusions}
\label{Sec:Conclusions}

This work has focused on an analysis of the redshift of a collection of photons illuminating a collapsing cloud and received by an asymptotic observer. The incident radiation is generated with arbitrary angular momenta by sources far away from the collapsing cloud, which for convenience are taken to be uniformly distributed. Moreover, this radiation does not interact with matter, but propagates freely, and eventually, part of it is detected by an asymptotic observer. Our results confirm the hypothesis put forward in paper~I (Ref.~\cite{nOoStZ14}), namely that the redshift can be considered as a tool capable of differentiating whether the outcome of a complete gravitational collapse is a naked singularity or a black hole. This confirmation consists in recovering the results in~I for the case of radial photons, but fortifies the proposal by demonstrating that the redshift of certain classes of nonradial photons exhibits similar behavior as the radial ones. Our results show that photons in the incident radiation whose directions (as specified relative to the asymptotic observer's frame) are close to the radial direction exhibit the same behavior as the radial ones. Specifically, there is a smooth decay to zero of the frequency shift for the case of event horizon formation, whereas in the naked singularity case the decay of the frequency shift is slower initially but shows an abrupt cutoff to zero after the observer crosses the Cauchy horizon.

Beyond these new insights offered by the redshift of nonradial photons, the inclusion of angular momenta added a new element in the proposal in~I. This new element consists in the structure and the properties of the shadow that the collapsing cloud casts into the sky of an asymptotic observer. Although the final state of the shadow turns out to be the same in both cases, we have found that the time scale required for the shadow to occult the source from the eyes of the asymptotic observer depends sensitively on whether the outcome of the collapse is a black hole or a naked singularity.

Summing up, in this work we have presented new evidence to support the view in~I that the structure of the redshifted radiation combined with properties of the shadow can be a tool in checking the validity of the weak cosmic censorship conjecture.

The proposal put forward in~I and tested further in this work is not the only proposal that has been considered in the literature. An alternative proposal is based on strong gravitational lensing effects. For instance, in Refs.~\cite{kVdNsC98,kVgE02,kVcK08} the gravitational lensing of sources due to a static, spherically symmetric solution of the Einstein-massless scalar field equations are compared to those of a Schwarzschild black hole. However, the choice of the nakedly singular spacetimes made in these works may be questioned on the grounds that no physical mechanism is known for the formation of these singular spacetimes starting from a regular set of initial data.  In Refs.~\cite{aV00,kHkM09}, the effects of a naked Kerr or Kerr-Newman ring singularity are compared to those of a Kerr or Kerr-Newman black hole. Although one might think that such singularities could be created by throwing highly spinning particles into a nearly extremal Kerr black hole, all the results so far suggest that this scenario does not work due to spin coupling effects~\cite{rW74,wH81,kDiS13,kD14} or backreaction effects (see Ref.~\cite{pZiVePrH13} and references therein). In that sense, the above mentioned alternative applications to test the weak cosmic censorship lack physical relevance.

One of the positive aspects of the proposal in~I is that the choice of the Tolman-Bondi model as a background spacetime avoids the above mentioned difficulties. Whenever the collapsing cloud is considered to be of finite spatial extent, an exterior Schwarzschild region joins smoothly across the cloud's surface. Such spacetimes can be generated by a regular set of initial data (for a discussion and properties of such data, see~\cite{nOoS11,nO12}), and the maximal future analytic extension of the initial data surface can be determined. Moreover, by choosing  the initial data appropriately, we obtain spacetimes with prescribed causal structure. This property combined with the fact that the metric in the maximal future analytic extension of the initial data surface is known, allows one to perform an analysis of the behavior of causal geodesics through the collapsing cloud. Two recent works~\cite{lKdMcB13b} and~\cite{kNnKhI03} also employ a Tolman-Bondi background to test the validity of the weak cosmic censorship or to study the shadow of the singularity. In the first work, the consideration of the authors were restricted to marginally bound collapse. Unlike our scenario, they considered a situation in which the photons are emitted from dust particles inside the cloud. In contrast to our results, they did not find any specific signature in the emitted radiation that can differentiate between the formation of a naked singularity and the formation of a black hole. The work in~\cite{kNnKhI03} treats only the globally naked self-similar Tolman-Bondi collapse and considers the impact of nonradial null geodesics escaping from the central singularity on the shadow as the central question. Although their approach has some resemblance to the present work, they have assumed that the photons originating from the central singularity have unbound frequencies so that they are received with nonzero frequency by the asymptotic observer. Their choice of infinite frequencies invokes plausibility arguments based on the unknown theory of quantum gravity, and their conclusions differ from those obtained in the present work.

It is interesting to mention briefly that a few conclusions reported in this work may extend beyond the family of the Tolman-Bondi spacetimes. For instance, it is well known since the work of Ames and Thorne~\cite{wAkT68} that the shadow arising during the formation of a black hole depends crucially upon the existence of the photosphere of the Schwarzschild exterior and has a weak dependence upon the details of the collapse. However, as far as the shadow of a naked singularity is concerned, many issues need to be considered in detail. For instance, it is not clear whether the existence of future directed null geodesics emanating from the central singularity is a generic property of null singularities or if it is associated with a particular feature of the Tolman-Bondi metric. If the first possibility occurs, then maybe some of the detail characteristics of the shadow can be extended to more general collapse models provided we have regular initial data whose evolution leads to a globally nakedly singular development.
It is needless to say that these issues need to be checked much more thoroughly.

What are the prospects that the proposal in~I in combination with the results of this work, can be called to decide between the two competing alternatives in a realistic collapse event?

As it has become clear from the present work, the crossing time $t_c := GM/c^3 = 10^{-5}(M/M_\odot)$s with $M_\odot = 2\times 10^{33}$g the solar mass is the relevant time scale. Since typical stellar masses vary in the range $(1-100)M_\odot$, the corresponding crossing time $t_c$ is of the order of a few $(10^{-3} - 10^{-5})$s which seems to be too short to be resolved with existing technology. However, cosmological scenarios offer better testing grounds. It seems by now that there is a consensus that supermassive black holes with masses in the range between $10^{5}M_\odot$ and $10^9 M_\odot$ are the objects that energize quasars and active galactic nuclei. Even though this hypothesis resolves one problem, the same hypothesis raises the delicate question of the mechanism (or mechanisms) generating their enormous masses (for an overview of this delicate problem and further references see~\cite{jJdWhLdH13}). A currently popular scenario regarding their formation is based on the old idea of supermassive stars proposed long ago by Hoyle and Fowler~\cite{fHwF63a,fHwF63b} (for an introduction of these configurations see Ref.~\cite{ShapiroTeukolsky-Book}). In this scenario, at the early Universe, i.e. at times corresponding to cosmological redshifts $5\leq z \leq 20$, the dark halos led to the formation of supermassive stars with masses between $10^5M_\odot$ and $10^6M_\odot$, and these supermassive stars collapse gravitationally to form a black hole. These seed black holes subsequently grew via the accretion process to reach the range $(10^5-10^9)M_\odot$. Although many assumptions enter in this scenario, it is important to realized that the validity of the cosmic censor conjecture plays a central role. Suppose, for example, that the outcome of the gravitational collapse of a supermassive dark star is a naked singularity instead of a black hole. In this event, the resulting shadow offers the best case scenario to be resolved by terrestrial instruments. However, the fact that the event takes place at a high cosmological redshift adds technicalities regarding the observability of such a collapsing event at least within the current technological capabilities. On the other hand, the analysis presented in this work stands on a firm basis and at the theoretical level differentiates between the two possibilities.


\acknowledgments
N. O. especially wants to thank Jonah Miller for his valuable assistance on high performance computing and display techniques. O.S. wishes to thank the Perimeter Institute for Theoretical Physics for hospitality. T.Z. thanks the Department of Physics at Queen's University and Kayll Lake for the hospitality during a sabbatical year. We  thank  Kayll Lake for discussion on some of the issues raised in this work.

This research was supported in part by CONACyT Grants No. 232390 and No. 234571, by a CIC Grant to Universidad Michoacana and by Perimeter Institute for Theoretical Physics. Research at Perimeter Institute is supported by the Government of Canada through Industry Canada and by the Province of Ontario through the Ministry of Research and Innovation.

\appendix
\section{Proof of Theorem~\ref{Thm:Stability}}
\label{App:Proof}

The proof is based on standard arguments from the theory of ordinary differential equations; see for instance Refs.~\cite{Walter-Book,BrauerNohel-Book}. Without loss of generality we assume that $x^*=0$. Since $U$ is open there exists $\delta > 0$ such that the ball $B_\delta(0)$ of radius $\delta$ centered at the origin is entirely contained in $U$: $B_\delta(0)\subset U$. Let $\varepsilon\in (0,\delta)$, and let $s_0 > 0$ and $x_0\in B_\varepsilon(0)$. Let $x(s)$ denote the maximally extended solution of Eq.~(\ref{Eq:PertDyn}), and define
$$
s_0^* := \sup\{ s\geq s_0 : x(s')\in B_\delta(0) \hbox{ for all $s_0\leq s' < s$} \}.
$$
We claim that $s_0^* = +\infty$ provided that $\delta > \varepsilon > 0$ are chosen small enough and $s_0 > 0$ is large enough, and that in this case $\lim\limits_{s\to \infty} x(s) = 0$. The statement of the theorem then follows by choosing $s_1$ large enough and $V := B_\varepsilon(0)$ with $\varepsilon > 0$ small enough.

To prove the claim, we decompose
$$
X(x) = A x + g(x),
$$
with $A := DX(0)$ and $g(x) := X(x) - A x$. According to Duhamel's principle, $x(s)$ satisfies
\begin{equation}
x(s) = e^{A(s-s_0)} x_0 + \int\limits_{s_0}^s e^{A(s-s')}
\left[ g(x(s')) + \frac{1}{s'} Y(s',x(s')) \right] ds',
\label{Eq:Duhamel}
\end{equation}
for all $s_0\leq s \leq s_0^*$. Since all the eigenvalues of $A$ have a negative real part, there exist positive constants $C > 0$ and $\beta > 0$ such that
$$
| e^{A t} | \leq C e^{-\beta t}
$$
for all $t\geq 0$. Furthermore, by the differentiability of $X(x)$ we can choose $\delta > 0$ small enough such that
$$
\frac{|g(y)|}{|y|} \leq \frac{\beta}{2C}
$$
for all $y\in B_\delta(0)\setminus \{ 0 \}$. Further, since $Y$ is bounded, there is a constant $K > 0$ such that
$$
|Y(s,y)| \leq \frac{\beta K}{2C}
$$
for all $s\geq s_0$ and all $y\in B_\delta(0)$. It then follows from Eq.~(\ref{Eq:Duhamel}) that
$$
|x(s)| \leq C e^{-\beta(s-s_0)} |x_0| 
 + \int\limits_{s_0}^s C e^{-\beta(s-s')} \left( \frac{\beta}{2C} |x(s')| + \frac{\beta K}{2C s'} \right) ds'
$$
for all $s_0\leq s \leq s_0^*$. Hence, the function $\phi(s) := e^{\beta(s-s_0)} |x(s)|$ satisfies
$$
\phi(s) \leq C\varepsilon 
 + \frac{\beta}{2}\int\limits_{s_0}^s  \left( \phi(s') + \frac{K}{s'} e^{\beta(s'-s_0)} \right) ds'
  =: \psi(s),\qquad s_0\leq s \leq s_0^*.
$$
Since
$$
\frac{d}{ds}\psi(s) = \frac{\beta}{2}\left( \phi(s) + \frac{K}{s} e^{\beta(s-s_0)} \right)
 \leq \frac{\beta}{2}\left( \psi(s) + \frac{K}{s} e^{\beta(s-s_0)} \right),
$$
we obtain
$$
\frac{d}{ds}\left[ e^{-\frac{\beta}{2}(s-s_0)}\psi(s) \right]
 \leq \frac{\beta}{2}\frac{K}{s} e^{\frac{\beta}{2}(s-s_0)}.
$$
Integrating and using $\psi(s_0) = C\varepsilon$ and $|x(s)| = e^{-\beta(s-s_0)}\phi(s) \leq e^{-\beta(s-s_0)}\psi(s)$ finally yields
\begin{equation}
|x(s)| \leq C\varepsilon e^{-\frac{\beta}{2}(s-s_0)} + \frac{\beta K}{2}\int\limits_{s_0}^s
 \frac{1}{s'} e^{-\frac{\beta}{2}(s-s')} ds'
\label{Eq:xEstimate}
\end{equation}
for all $s_0\leq s\leq s_0^*$. The right-hand side is bounded from above by the constant $C\varepsilon + K/s_0$. Therefore, if we choose $\varepsilon\in (0,\delta)$ small enough and $s_0 > 0$ large enough such that $C\varepsilon + K/s_0 < \delta$, it follows that the solution $x(s)\in B_\delta(0)$ stays forever inside the ball $B_\delta(0)$ and thus exists for arbitrarily large times $s\geq s_0$.

In order to conclude the proof, we need to show that $|x(s)|$ converges to zero when $s\to\infty$. For this, we first note that the first term on the right-hand side of Eq.~(\ref{Eq:xEstimate}) vanishes for $s\to\infty$ since it is exponentially damped. As for the second term, using L'H\^opital's rule we find
$$
\lim\limits_{s\to \infty}\frac{\beta}{2}\int\limits_{s_0}^s \frac{1}{s'} e^{-\frac{\beta}{2}(s-s')} ds'
 = \lim\limits_{s\to \infty} \frac{\int\limits_{s_0}^s \frac{1}{s'} e^{\frac{\beta}{2} s'} ds'}
 {\frac{2}{\beta} e^{\frac{\beta}{2} s}}
 = \lim\limits_{s\to \infty} \frac{\frac{1}{s} e^{\frac{\beta}{2} s}}{e^{\frac{\beta}{2} s} } = 0.
$$
This concludes the proof of the theorem.

\section{Appearance of the naked singularity in the self-similar case}
\label{App:SS}

In this appendix, we compute the critical angle of incidence $\hat{\alpha}$ for the particular case of a free-falling comoving observer in a nakedly singular, self-similar Tolman-Bondi spacetime. In this case, the presence of the homothetic Killing vector field $\xi$ associated with the self-similarity allows one to analytically integrate the null geodesic flow. For a complete analysis for the behavior of the null geodesics in these spacetimes, we refer the reader to our recent work in~\cite{nOoStZ15b}. The analysis and notation used in this appendix are based on that work. For related numerical work we refer the reader to Ref.~\cite{kNnKhI03}.

\begin{figure}[h!]
\begin{center}
\includegraphics[width=8.0cm]{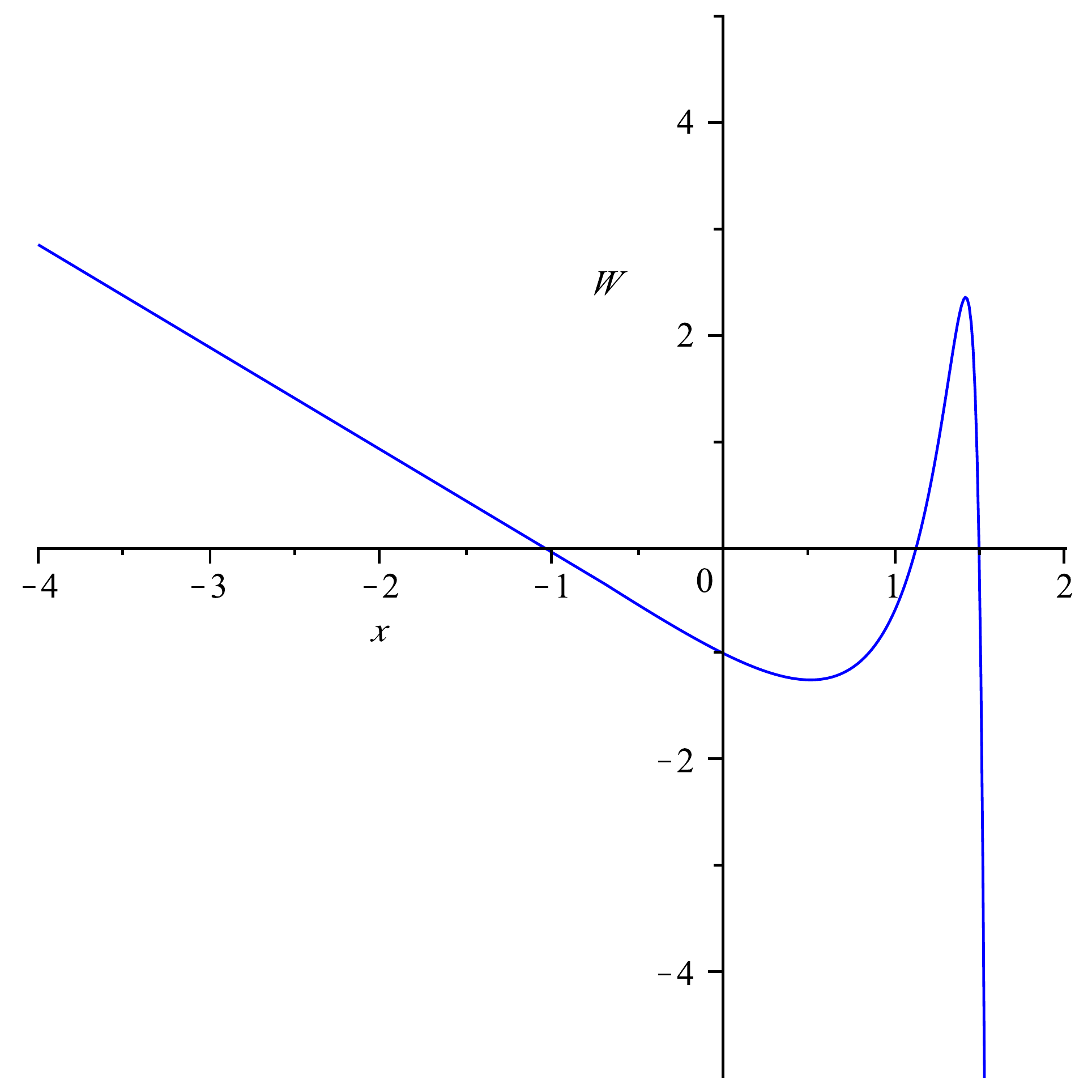}
\end{center}
\caption{\label{Fig:EffPot} A plot of the effective potential $W(x)$ for the parameter value $\lambda = 0.6$.}
\end{figure}

For a self-similar collapsing spacetime, the metric coefficients in Eq.~(\ref{Eq:Metric}) are given by
\begin{equation}
e^{\Phi} = 1,\qquad
e^{\Psi} = F(x),\qquad
r = R S(x),\qquad x := \frac{t}{R} < \frac{1}{\lambda}
\label{Eq:MetricCoeffSS}
\end{equation}
with the functions
$$
F(x) := \frac {1-\frac {\lambda x}{3}}{(1-\lambda x)^{1/3}},\quad 
S(x) := (1-\lambda x)^{2/3},\qquad x < 1/\lambda.
$$
Here, $\lambda > 0$ is a positive parameter determining the compactness ratio
$$
\frac{2m(R)}{R} = \frac{4}{9}\lambda^2
$$
at time $t=0$. Note that in this model, this ratio is independent of $R$. Since $r' = F > 0$, there are no shell-crossing singularities in this model; however, there is a shell-focusing singularity at $x = 1/\lambda$. When $\lambda < \lambda^* \simeq 0.638$, this shell-focusing singularity is naked. The causal structure of the resulting spacetime has been studied a long time ago; see for instance Refs.~\cite{mCaT71,kLtZ90,aOtP87,dC94,pB95,bNfM02,bCaC05}. Recently~\cite{nOoStZ15b}, we provided a full qualitative picture for the null geodesic flow (with and without angular momentum). Our analysis makes use of the homothetic Killing vector field $\xi$, and the main features of this flow depend on the effective potential
$$
W(x) := \frac{x^2 - F^2(x)}{S^2(x)},\qquad x < \frac{1}{\lambda},
$$
see Fig.~\ref{Fig:EffPot} for a plot of this function for the parameter value $\lambda = 0.6$. As can be seen from this plot (cf. Lemma~2 in~\cite{nOoStZ15b}), $W$ has three simple zeros located at $x = J_a$, $a=0,1,2$, the values $J_a$ satisfying $J_0 < 0 < J_1 < J_2 < 1/\lambda$. Geometrically, these zeros correspond to the set of points where the homothetic Killing vector field is null. $W$ is positive for $x < J_0$ and $J_1 < x < J_2$ and negative for $J_0 < x < J_1$ and $J_2 < x < 1/\lambda$, and it has a local maximum at some point $x = x_c\in (J_1,J_2)$.

As a consequence of the conservation of angular momentum $\ell$ and the presence of the conserved quantity $C$ associated with the homothetic Killing vector field, the motion in the $x$ direction is restricted to the set $W(x) \leq 1/\beta^2$, with $\beta := \ell/C$ an ``impact parameter". For the following, the critical impact parameter $\beta_c := 1/\sqrt{W(x_c)}$ corresponding to the local maximum of $W$ plays an important role. The main qualitative features of the null geodesic flow which will be used here can be summarized as follows (cf. Figs.~2 and 4 in~\cite{nOoStZ15b}):
\begin{enumerate}
\item[(i)] The surface $x = J_0$ is generated by ingoing radial null geodesics terminating at the central singularity $(\tau,R) = (0,0)$, while the surfaces $x = J_1$ and $x = J_2$ are spanned by outgoing radial null geodesics emanating from the central singularity. The surface $x = J_1$ describes the Cauchy horizon, and it is generated by the earliest radial outgoing null geodesics emanating from the singularity.
\item[(ii)] For small impact parameters, $|\beta| < \beta_c$, all the outgoing null geodesics registered by a free-falling observer at constant $R$ originate from the far past when $x < J_1$ and from the central singularity when $x > J_1$.
\item[(iii)] For large impact parameters, $|\beta| > \beta_c$, a free-falling observer at constant $R$ ``sees" the following: all the outgoing null geodesics registered by the observer emanate from the far past when $x < x_1(\beta)$, no null geodesics are registered when $x_1(\beta) < x < x_2(\beta)$ and all the outgoing null geodesics registered by the observer emanate from the central singularity when $x > x_2(\beta)$. Here, $x_1(\beta) < x_2(\beta)$ refer to the positive roots of $W(x) = 1/\beta^2$.
\end{enumerate}

The impact parameter $\beta$ is related to the angle of incidence $\alpha$ defined in Eq.~(\ref{Eq:Defalpha}) in the following way:
\begin{equation}
\cos\alpha = -\frac{F(x)\pm x Q_\beta(x)}{x \pm F(x) Q_\beta(x)},\qquad
\sin\alpha = -\frac{\beta}{S(x)} \frac{x^2 - F^2(x)}{x \pm F(x)Q_\beta(x)},
\label{Eq:alphabeta}
\end{equation}
where $Q_\beta(x) := \sqrt{1 - \beta^2 W(x)}$. In the radial case, $\beta = 0$ and thus $\cos\alpha = \mp 1$ and $\sin\alpha = 0$, showing that in this case the upper sign corresponds to outgoing null geodesics, such that $\alpha = \pi$. In order to determine the appearance of the naked singularity as seen by a free-falling observer at constant $R$, we have to find the critical angle $\hat{\alpha}$ such that all the outgoing null geodesics registered by the observer with $\alpha\in (\hat{\alpha},2\pi - \hat{\alpha})$ emanate from the central singularity and all the other outgoing geodesics originate from the far past (see Sec.~\ref{Sec:Step2}). Based on the properties (i)--(iii) listed above we find the following result for $\hat{\alpha}$.

\begin{figure}[h!]
\begin{center}
\includegraphics[width=8.0cm]{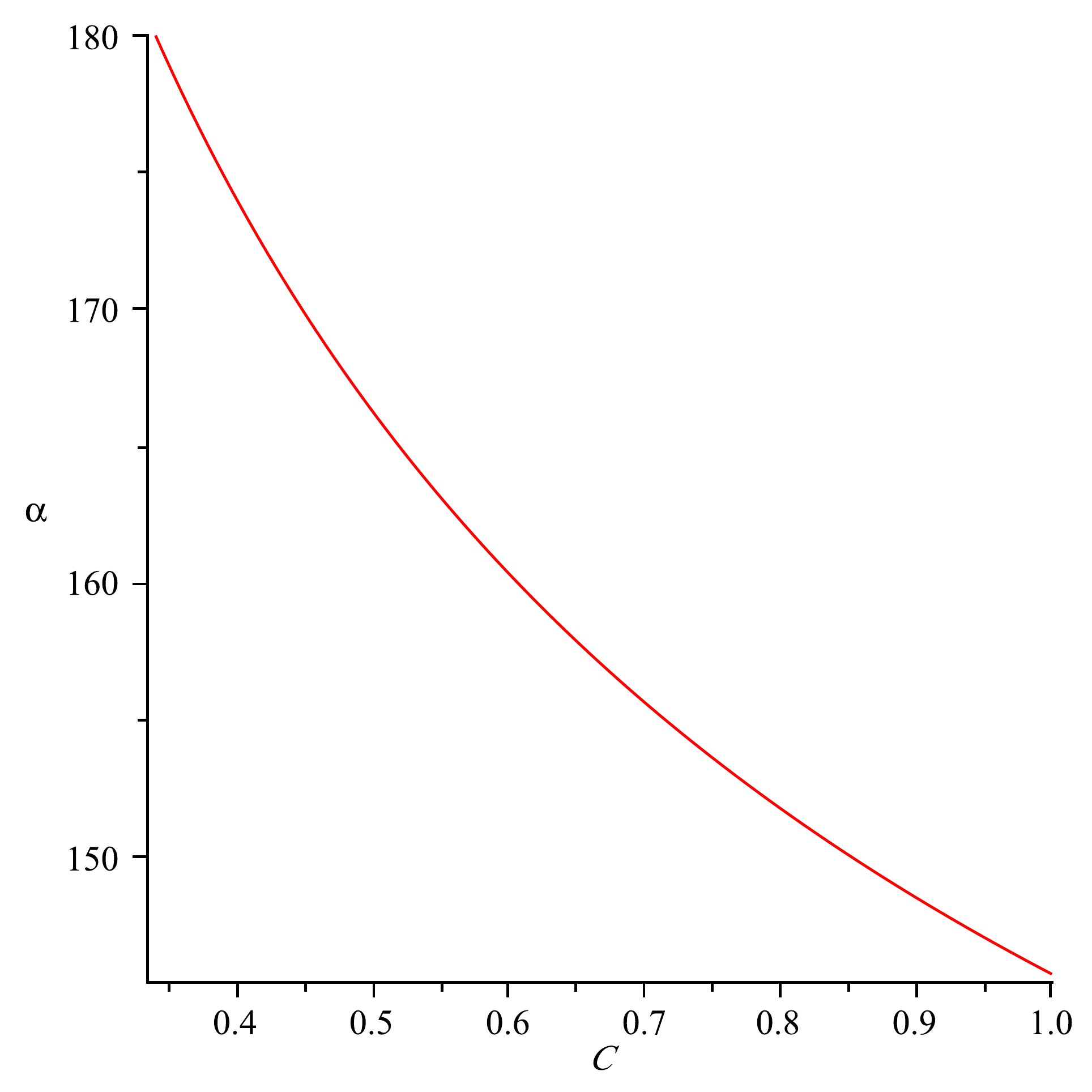}
\end{center}
\caption{\label{Fig:Appearance} The critical angle of incidence $\hat{\alpha}$ versus the compactness ratio ${\cal C} := 2m/r$ along the world line of a free-falling observer at constant $R$. The interval shown corresponds to the interval $J_1 < x < J_{AH}$. As in the previous figure, the parameter $\lambda$ is set to $0.6$. We note that the behavior of $\hat\alpha$ in the self-similar case resembles the behavior of $\hat\alpha$ in the bound collapse case, see the central and right panels of Fig.~\ref{Fig:Alphas}.}
\end{figure}

First, when $x < J_1$ there are no null geodesics emanating from the singularity, since $x = J_1$ describes the Cauchy horizon. Consequently, $\hat{\alpha} = \pi$ for $x < J_1$. Next, consider a fixed value of $x$ in the interval $J_1\leq x < x_c$. In this case, the impact parameter $\beta$ is restricted by the inequality $\beta^2 \leq 1/W(x)$. However, only those null geodesics which have $\beta^2 < \beta_c^2$ and correspond to the upper sign in Eq.~(\ref{Eq:alphabeta}) emanate from the central singularity (see Fig.~4 in~\cite{nOoStZ15b}). Hence, the angles of incidence $\alpha$ registered by the observer corresponding to null geodesics originating from the singularity are those for which
$$
-1 \leq \cos\alpha < -\frac{F(x) + xQ_{\beta_c}(x)}{x + F(x) Q_{\beta_c}(x)},\qquad
J_1 \leq x < x_c.
$$
Next, assume $x$ lies in the interval $x_c\leq x \leq J_2$. As in the previous case, the null geodesics with impact parameter $\beta^2 < \beta_c^2$ corresponding to the upper sign in Eq.~(\ref{Eq:alphabeta}) emanate from the singularity. However, when $\beta^2 > \beta_c^2$, the null geodesics corresponding to both signs in Eq.~(\ref{Eq:alphabeta}) are also registered by the observer as outgoing, and they both emanate from the central singularity (see the right panel of Fig.~4 in~\cite{nOoStZ15b}). Therefore, $\cos\alpha$ varies over all those values on the right-hand side of Eq.~(\ref{Eq:alphabeta}) corresponding to the upper sign for which $\beta^2 \leq 1/W(x)$ [$1\geq Q_\beta(x)\geq 0$] and all those values corresponding to the lower sign for which $\beta_c^2 <\beta^2 \leq 1/W(x)$ [$Q_{\beta_c}(x) > Q_\beta(x)\geq 0$]. Therefore, the angles $\alpha$ corresponding to null geodesics originating from the singularity are those for which
$$
-1 \leq \cos\alpha < -\frac{F(x) - xQ_{\beta_c}(x)}{x - F(x) Q_{\beta_c}(x)},\qquad
x_c\leq x < J_2.
$$
Finally, when $x$ lies in the interval $J_2 < x < 1/\lambda$, $W(x)$ is negative, and hence $\beta$ can take any real value. However, as in the previous case, only those null geodesics corresponding to the upper sign in Eq.~(\ref{Eq:alphabeta}) or those with the lower sign and $\beta^2 > \beta_c^2$ emanate from the singularity. Summarizing, we thus have
\begin{equation}
\hat{\alpha}(x) = \left\{ \begin{array}{ll}
\pi,  & x < J_1,\\
\arccos\left( -\frac{F(x) + xQ_{\beta_c}(x)}{x + F(x) Q_{\beta_c}(x)} \right), 
& J_1\leq x < x_c,\\
\arccos\left(  -\frac{F(x) - xQ_{\beta_c}(x)}{x - F(x) Q_{\beta_c}(x)} \right), 
& x_c\leq x < \frac{1}{\lambda}.
\end{array} \right.
\end{equation}
Notice that $\hat{\alpha}$ is continuous at $x = J_1$ and $x_c$. Moreover, it has a well-defined limit as $x\to J_2$, since
$$
\lim\limits_{\substack{x\to J_2 \\ x < J_2}}
 -\frac{F(x) - xQ_{\beta_c}(x)}{x - F(x) Q_{\beta_c}(x)}
= \frac{1 - \beta_c^2\frac{J_2^2}{S^2(J_2)}}{1 + \beta_c^2\frac{J_2^2}{S^2(J_2)}}.
$$
The critical angle $\hat{\alpha}$ as a function of the compactness ratio along the trajectory of a free-falling observer at constant $R$ is shown in Fig.~\ref{Fig:Appearance} for the parameter value $\lambda = 0.6$.

Although in the parametrization given in Eq.~(\ref{Eq:MetricCoeffSS}) the singularity already appears in the $t=0$ slice, one can show that the spacetime has a regular center at $R = 0$ for negative values of $t$. Furthermore, it is possible to match the self-similar spacetime described by Eq.~(\ref{Eq:MetricCoeffSS}) to a Schwarzschild vacuum exterior spacetime. This can be accomplished at a constant $R$ hypersurface by satisfying the standard junction conditions; see Ref.~\cite{kL00} and references therein. The apparent horizon in the resulting spacetime is determined by the equation $2m = r$, and it coincides with the event horizon in the Schwarzschild exterior, while in the interior it is described by the constant $x = J_{AH}$ surface given by $S(x) = 4\lambda^2/9$, that is,
$$
J_{AH} = \frac{1}{\lambda}\left[ 1 - \left( \frac{2\lambda}{3} \right)^3 \right].
$$
Since $F(J_{AH}) - J_{AH} = 4\lambda^2/9 > 0$, it follows that $J_{AH} > J_2$ and that $x = J_{AH}$ is a spacelike surface. For $\lambda = 0.6$ we obtain the numerical values
$$
J_0 \simeq -1.027,\quad
J_1 \simeq 1.129,\quad
J_2 \simeq 1.495,\quad
J_{AH} \simeq 1.560. 
$$

\bibliographystyle{unsrt}
\bibliography{../References/refs_collapse}

\end{document}